\documentclass[a4paper,11pt]{article}
\pdfoutput=1
\usepackage{jheppub} 
\usepackage{graphicx,color,rotating}
\usepackage{hyperref}
\usepackage{epsfig,color}
\usepackage{slashed}
\usepackage{amsfonts}
\usepackage{ulem}

\usepackage{tabularx}
\usepackage{graphicx}
\usepackage{adjustbox}
\usepackage{tabularx}

\begin{document}
\title{Iso-doublet Vector Leptoquark solution to  the Muon $g-2$, $R_{K, K^*}$, $R_{D,D^*}$,
  and $W$-mass Anomalies}

\renewcommand{\thefootnote}{\arabic{footnote}}

\author{
Kingman Cheung$^{1,2}$, Wai-Yee Keung$^{3}$,  and Po-Yan Tseng$^{1}$}
\affiliation{
$^1$ Department of Physics and CTC, National Tsing Hua University,
Hsinchu 300, Taiwan \\
$^2$ Division of Quantum Phases and Devices, School of Physics, 
Konkuk University, Seoul 143-701, Republic of Korea \\
$^3$ Department of Physics, University of Illinois at Chicago,
Illinois 60607 USA \\
}

\date{\today}

\abstract{
  We investigate the iso-doublet vector leptoquark $V_2$ as a solution to the
  $B$ anomalies $R_{K, K^*}$ and $R_{D,D^*}$, as well as explaining the muon and
  electron anomalous magnetic moments, and the very recent $W$-mass anomaly.
}

\maketitle

\section{Introduction}
Recent trends in hunting for new physics beyond the standard model (SM) can divided into
(i) high energy frontier, (ii) precision frontier, and (iii) cosmology frontier.  While
the high energy frontier has not been finding anything new other than the
discovery of the Higgs boson and the cosmology frontier involves large uncertainties
associated with observations, the precision frontier, on the other hand, has shown
some surprising results. Namely, there are a number of anomalies in $B$ meson
decays, the muon anomalous moment, and the very recent $W$-boson mass
measurement \cite{CDF:2022hxs}.

After accumulating data for a number of years, there exist persistent discrepancies
between the SM predictions and the experimental results for the flavor-changing
neutral current rare decays of $B$ mesons in $b \to s \ell\ell$. In particular, the
lepton-flavor universality violation in $B\to K$ transition observed by LHCb
\begin{equation}
  R_K = \frac{ {\rm BR}( B \to K \mu^+ \mu^-) }{ {\rm BR}( B \to K e^+ e^-) } \;, \,\;\;\;
  R_{K*} = \frac{ {\rm BR}( B \to K^* \mu^+ \mu^-) }{ {\rm BR}( B \to K^* e^+ e^-) }
  \;,
\end{equation}
with the measurements \cite{LHCb:2021trn,LHCb:2017avl}
\begin{equation}
  R_K = 0.846\,^{+0.042}_{-0.039}\, ^{+0.013}_{-0.012} \;,  \;\;\;
  \mbox{for 1.1 GeV$^2  < q^2 < 6$ GeV$^2$} \;,
\end{equation}
\begin{equation}
  R_{K^*} = \left \{ \begin{array}{lr}
  0.66 \, ^{+0.11}_{-0.07} \, \pm 0.03 & 0.045 \;{\rm GeV}^2 < q^2 < 1.1 \; {\rm GeV}^2\,,\\
  0.69 \, ^{+0.11}_{-0.07} \, \pm 0.05 &  1.1 \;{\rm GeV}^2 < q^2 < 6.0\; {\rm GeV}^2\,,
  \end{array}
   \right .
\end{equation}
which deviate from the SM predictions by as large as $3\sigma$. 
The advantage of using ratios is that the ratio can have a lot of hadronic uncertainties
in each branching ratio measurement to be eliminated.
Precise measurements of these ratios with significant deviations from the SM
predictions can hint at new physics.  The same short distance process $b\to s\ell\ell$
is also responsible for $B_s^0 \to \ell^+ \ell^-$.  

Another set of observables is related to the short-distance process $b \to c \ell\nu$ and
the observables are \cite{HFLAV:2019otj}
\begin{eqnarray}
R_{D} &=& \frac{ {\rm BR}(B \to D \tau \nu) } { {\rm BR}(B \to D \ell \nu) } =
0.340 \pm 0.027 \pm 0.013\,,  \nonumber \\
R_{D^*} &=& \frac{ {\rm BR}(B \to D^* \tau \nu) } { {\rm BR}(B \to D^* \ell \nu) } =
0.295 \pm 0.011 \pm 0.008 \;, 
\end{eqnarray}
and the combined discrepancy to SM prediction is at the $3.1\sigma$ level.

Another long-standing experimental anomaly is the muon anomalous moment
(aka. $g-2$).  The most recent muon $g - 2$ measurement was performed by the 
E989 experiment at Fermilab, which  reported the new result \cite{Muong-2:2021ojo}
\begin{equation}
  \label{amu}
\Delta a_\mu = (25.1 \pm 5.9) \times 10^{-10} \;,
\end{equation}
which deviates at the level of $4.2\sigma$ from the most recent SM prediction.
\footnote{
  There was a recent lattice calculation of the eading hadronic contribution to the
  muon magnetic moment \cite{Borsanyi:2020mff}, which brought the SM prediction
  within $1\sigma$ of the experimental result. Yet,  one has to wait further for the
  lattice community to settle on the calculation.  We will focus on Eq.~(\ref{amu})
  for the deviation of muon $g-2$.
}
On the other hand, the $g-2$ measurements for electron also show discrepancy
with the SM prediction.  The electron $g-2$ was used to be the most precise determination
of the fine-structure constant $\alpha$.  Nevertheless, there were two contradicting
determinations of $\alpha$ \cite{Aoyama:2012wj,Aoyama:2019ryr} resulting 
in two theory predictions, which deviate from the experimental measurement 
\cite{Hanneke:2008tm} in opposite direction, given by
\begin{eqnarray}
  \Delta a_e^{\rm LKB} &=a^{\rm exp}_e - a^{\rm LKB}_e = (4.8 \pm 3.0) \times 10^{-13}
  \;, \nonumber \\
   \Delta a_e^{\rm B} &=a^{\rm exp}_e - a^{\rm B}_e = (-8.8 \pm 3.6) \times 10^{-13}
   \;.
\end{eqnarray}
In the following analysis, we show the results for both cases of $\Delta a_e$.
Note that if we want to explain the $\Delta a_\mu$ with new physics, it is 
very likely subject to constraints coming from lepton-flavor-violating (LFV) decays,
such as $\mu \to e \gamma$.

There has been a vast literature in explaining all or part of the anomalies. Especially,
the leptoquark (LQ) provides viable explanations for the $R_{K,K^*}$ and/or
$R_{D,D^*}$ anomalies.  It was illustrated in Ref.~\cite{Angelescu:2021lln} that
only the iso-singlet vector LQ $U_1$ with the SM quantum numbers
$({\bf 3}, {\bf 1}, 2/3 )$ can explain both $R_{K,K^*}$ and $R_{D,D^*}$,
while the scalar LQ's $S_1,S_3, R_2$
and the iso-triplet vector LQ $U_3\, ({\bf 3}, {\bf 1}, 2/3 )$ can only explain one of
the anomalies.  There has been very few studies on the isospin-doublet vector LQ,
denoted by $V_2 \,( {\bf 3}, {\bf 2}, 5/6 )$. 
In this work, we attempt to fill the gap by showing that the iso-doublet vector LQ
$V_2$ can satisfy the anomalies $R_{K,K^*}$ and $R_{D,D^*}$, as well as it can
satisfy the $\Delta a_\mu$ and $\Delta a_e$ subject to LFV constraints.

Very recently, there was a new $W$-boson mass measurement by the CDF
Collaboration \cite{CDF:2022hxs}:
\[
M_W = 80.4335 \pm 0.0094 \; {\rm GeV}
\]
which is about $7\sigma$ above the SM prediction of
$M_{W,SM} = 80.361 \pm 0.006$ GeV \cite{ParticleDataGroup:2016lqr}.
Additional corrections to the $W$-boson mass can come from
new physics, which can be conveniently parameterized in terms of the
Peskin-Takeuchi parameters $S,T,U$ \cite{Peskin:1991sw}.  The $W$-boson mass
correction is most sensitive to the $T$ parameter. It is well known that electroweak
doublet with a mass splitting between the upper and lower components contributes
positively to the $T$ parameter, thus gives a positive contribution to the $W$ mass.

We highlight the capabilities of the LQ $V_2$ in solving the anomalies:
\begin{enumerate}
\item The more popular explanation of $R_{K,K^*}$ anomaly is by decreasing the
  $b\to s \mu^+ \mu^-$ using the left-handed coupling to the muon, such that the
  combination of the Wilson coefficients $C_9 $ and $C_{10}$ is in the form
  $C_9 = - C_{10}$ in accordance with the best fit \cite{Altmannshofer:2021qrr}.
  This is the case of the iso-singlet vector LQ $U_1$,  On the other hand, 
  the iso-doublet vector LQ $V_2$ couples to the right-handed lepton. such that it
  contribute to $C_9$ and $C_{10}$ in the combination of $C_9 = C_{10}$. We therefore
  require $V_2$ to couple to electron and to increase $b \to s e^+ e^-$.

\item $V_2$ can enhance substantially the $\Delta a_\mu$ and change $\Delta a_e$ in
  both directions using different combinations of couplings. Nevertheless, in order to
  reach with $2\sigma$ of $\Delta a_\mu$ one of the couplings has to be very large.
  This is due to the constraint from the $\mu \to e \gamma$. 
  
\item $R_{D,D^*}$ can be easily satisfied.

\item The iso-doublet LQ $V_2$ can couple to the weak gauge bosons and thus modifies
  the $S,T,U$ parameters. It can explain the very recent $W$-mass anomaly measured
  by CDF. 
  
\end{enumerate}
  
The organization of this work is as follows. In the next section, we introduce the
interactions of the vector LQ. In Sec.~3, we describe the effects of the vector LQ
on various observables that we used in this analysis. In Sec.~4, we give the
results and the valid parameter space. We conclude in Sec.~5.

\section{Vector Leptoquark Interactions}

The weak iso-doublet vector LQ is denoted by $V_2$ with the SM quantum numbers
$( {\bf \bar 3},\,  {\bf 2},\, 5/6)$. The $V_2$ is written as
\[
V_2 = \left( \begin{array}{c} V^{+4/3} \\
  V^{+1/3} \end{array} \right )
\]
For simplicity we drop the subscript of $V_2$ from now on.
The gauge interactions of the $V_2$ is given by
\begin{eqnarray}
  {\cal L}_{V_2} &=&  -\frac{1}{2} V_{\mu\nu}^\dagger \, V^{\mu\nu}
  + M_V^2 V^{\dagger}_\mu \,   V^{\mu} \nonumber \\
  &+&
      i g_3 V^{\dagger}_\mu \frac{\lambda^A}{2} V_\nu \, G^{A,\mu\nu}
  + i g_2 V^{\dagger}_\mu \frac{\tau^k}{2} V_\nu W^{k,\mu\nu}
  + i g_1 V^\dagger_\mu Y  V_\nu B^{\mu\nu}
 \end{eqnarray}
where  $V_{\mu\nu} = \sum_{i=1,2}  D_\mu V^i_\nu - D_\nu V^i_\mu$ and
\begin{equation}
  D_\mu = \partial_\mu + i g_1 Y B_\mu + i g_2 \frac{\tau^k}{2} W^k_\mu
  + i g_3 \frac{\lambda^A}{2} G_\mu^A
\end{equation}
where the $SU(2)_L$ index $k = 1,2,3$ and the color index $A=1, ..., 8$.
\footnote{
  Here we have take all $\kappa=1$ and $\tilde{\kappa} = 0$
  as in the convention of Ref.~\cite{Altmannshofer:2020ywf} with the assumption
  $V_2$ coming from spontaneous breakdown of a gauge symmetry.
  Note the major difference is that $U_1$ discussed in Ref.~\cite{Altmannshofer:2020ywf}
  always connects down-type quarks and charged leptons of the same chiralities, while
  here the $V_2$ always connects them with opposite chiralities.
  }
In order to extract the electromagnetic interaction with $V_2$ we  transform $B_\mu $ and
$W^3_\mu$ into $A_\mu$ and $Z_\mu$ by the usual transformation:
\[
B_\mu = c_w A_\mu - s_w Z_\mu \,, \qquad  W^3_\mu = s_w A_\mu  + c_w Z_\mu \,,
\]
where $c_w= \cos \theta_W, \, s_w= \sin \theta_W$ and $\theta_W$ is the weak mixing angle.
As long as the electromagnetic interaction is concerned, we simplify the Lagrangian as
\begin{equation}
  {\cal L}_{V_2, em} =  - \frac{1}{2} V_{\mu\nu}^\dagger V^{\mu\nu}
  + i e Q_V V^\dagger_\mu V_\nu F^{\mu\nu} + ... 
\end{equation}
where $D_\mu = \partial_\mu + i e Q_V A_\mu + ... $. 
Extracting the triple vertex  $V^\dagger V A $ we obtain
\begin{eqnarray}
  {\cal L}_{V^\dagger V A} &=&  -\frac{1}{2} \left [
    i eQ_V ( \partial_\mu V_\nu^\dagger - \partial_\nu V^\dagger_\mu ) (A^\mu V^\nu - A^\nu V^\mu)
    -  i eQ_V (A_\mu V^\dagger_\nu - A_\nu V_\mu^\dagger) ( \partial^\mu V^\nu - \partial^\nu V^\mu)
    \right ] \nonumber \\
  &&
    +  ie Q_V V^\dagger_\mu V_\nu ( \partial^\mu A^\nu - \partial^\nu A^\mu ) \nonumber \\
    &=&
    -ie Q_V (\partial_\mu V_\nu^\dagger - \partial_\nu V_\mu^\dagger ) A^\mu V^\nu
    +ie Q_V  A_\mu V_\nu^\dagger (\partial^\mu V^\nu - \partial^\nu V^\mu )
    +ie Q_V  V^\dagger_\mu V_\nu (\partial^\mu A^\nu - \partial^\nu A^\mu ) \nonumber
\end{eqnarray}
We assign the 4-momenta and polarization vectors for $V^\dagger,\, V, A$ as
\[
V^\dagger:  p' ,\; \lambda'\,; \qquad
V:  p ,\; \lambda \,; \qquad
A: k,\; \epsilon \,.
\]
then we obtain the triple vertex as
\begin{equation}
  {\cal L}_{\rm V^\dagger V A} = e Q_V \left[
    g_{\alpha\beta} ( p - p')_\gamma + g_{\beta\gamma} ( p' - k )_\alpha
    + g_{\gamma\alpha} ( k -p )_\beta \right ]  \, \lambda^\alpha \lambda^{'\beta} \epsilon^\gamma
\end{equation}
Therefore, the interaction of $V$ with the photon is the same as the conventional
charged vector boson.

The interactions of $V_2$  with SM fermions are given by
\footnote{In principle, there could be diquark couplings to $V_2$ 
but they would lead to dangerous proton decay \cite{Greljo:2021npi}. 
Therefore, we do not include them here.}
\begin{eqnarray}
\label{eq_L_V2}
  {\cal L}_{Vff} &=&
   X^{RL}_{ij} \epsilon^{ab}\, \overline{d_R^{c,i}} \, \gamma^\mu V^a_\mu \,    L^{j,b}_L
   +X^{LR}_{ij} \epsilon^{ab}\, \overline{Q_L^{c,i,a}} \, \gamma^\mu V^b_\mu \,    e^{j}_R
   + {h.c.}   \nonumber \\
   &=& X^{RL}_{ij}  \left [  \overline{d_R^{c,i}} \,\gamma^\mu  \ell_L^j \, V^{+4/3}_\mu
      -  \overline{d_R^{c,i}} \,\gamma^\mu   \nu^j_L V_\mu^{+1/3} \right ] \nonumber \\
&&   + X^{LR}_{ij}  \left[
     \overline{u_L^{c,i}} \,\gamma^\mu  e_R^j \, V^{+1/3}_\mu
     -  \overline{d_L^{c,i}} \,\gamma^\mu  e_R^j \, V^{+4/3}_\mu \right ] + h.c.\,,
   \end{eqnarray}
where we assume that the down-type quarks and charged leptons are in the
   mass eigenstates while the CKM matrix is associated with the up-type quarks.
    To convert the up-type quarks of the interaction basis in Eq.(\ref{eq_L_V2}) into
    the mass eigenstates, we need to replace $\overline{u^{c,i}_L}$ in Eq.(\ref{eq_L_V2}) by $V_{ij}\overline{u^{c,i}_L}$  via CKM matrix $V_{ij}$. 


\section{Various Observables }
In this section, we summarize various observables that we are going to include
in this analysis, namely, the muon anomalous magnetic moment (aka $g-2$) and
lepton-flavor violating radiative decays, and $B$ anomalies $R_{K,K^*}$ and
$R_{D,D^*}$.  The $W$-boson mass is accounted for the mass splitting the isospin
components of the $V_2$, which has negligible effects on the low-energy observables.

\subsection{$g-2$ and $\ell_i \to \ell_j \gamma$ }
The amplitudes for the lepton anomalous magnetic moment $\Delta a_\ell$ and
lepton-flavor violating radiative decays $\ell_i \to \ell_j \gamma$ are related.
We can write the transition amplitude for 
\[
\ell_i (p) \to \ell_j (p-q) \; \gamma (q)
\]
where the 4-momentum $p$ is coming and $q$ is outgoing, as
\begin{eqnarray}
    \label{t1}
  T = e \epsilon^{\mu *}(q) m_{\ell_i} \,
  \bar u(p-q) i \sigma_{\mu\nu} q^\nu \, \left [
    A^{\ell_i \ell_j}_L P_L + A^{\ell_i\ell_j}_R P_R \right ] \, u(p)  \;.
\end{eqnarray}
Then the partial width of $\Gamma(\ell_i \to \ell_j \gamma)$ can be expressed as
\begin{equation}
  \Gamma(\ell_i \to \ell_j \gamma) = \frac{\alpha_{\rm em}}{4} m_{\ell_i}^5
  \left ( |A^{\ell_i \ell_j}_L|^2 + |A^{\ell_i \ell_j}_R|^2 \right )  \;,
\end{equation}
where the mass of the daughter lepton $\ell_j$ is ignored.

On the other hand, the anomalous magnetic moment form factor of the lepton $\ell$
is given by 
\begin{eqnarray}
  {\cal L}_{g-2} &=& \frac{e\, \Delta a_\ell} { 4 m_\ell } \bar \ell \sigma_{\mu\nu} \ell \, F^{\mu\nu}
  \nonumber \\
  \label{t2}
          &=& \frac{e\, \Delta a_\ell} { 2 m_\ell} \bar \ell i \sigma_{\mu\nu} (q^\nu )\,\ell  \, A^\mu 
\end{eqnarray}
where $\Delta a_\ell$ is the anomalous magnetic moment of the lepton and here
$q$ is coming into the vertex. Comparing
Eq.~(\ref{t1}) and Eq.~(\ref{t2}) we have 
\begin{equation}
 \Delta a_\ell  = - m_\ell^2 ( A^{\ell \ell}_L + A^{\ell \ell}_R  ) \;.
\end{equation}

We adopted the formulas here from Ref.~\cite{Hati:2019ufv}
  for $\ell_i \to \ell_j \gamma$
\begin{eqnarray}
  A^{\ell_i \ell_j}_L &=& - \frac{N_C} {16 \pi^2 M_V^2}\, \sum_k \biggr [
    - 2  X^{LR*}_{k \ell_j} X^{RL}_{k \ell_i} \frac{m_k}{m_{\ell_i}}\, (Q_V + Q_{b^c} ) \nonumber\\
    &&
    + \left( X^{LR*}_{k \ell_j} X^{LR}_{k \ell_i}   +  X^{RL*}_{k \ell_j} X^{RL}_{k\ell_i}
     \frac{m_{\ell_j}} {m_{\ell_i}} \right) \;
    \left( - \frac{5}{6} Q_V - \frac{2}{3} Q_{b^c}
 \right ) \biggr ]\,,\\
 A^{\ell_i \ell_j}_R &=& A^{\ell_i \ell_j} _L ( L \leftrightarrow R ) \nonumber
\end{eqnarray}
where $Q_{b^c} = - Q_b$ denotes the electric charge of the charge conjugate of the $b$ quark, and $Q_V = +4/3$ refers to the upper component of $V_2$.
We have made use of the fact that all down-type quark masses are much smaller
than the LQ mass $m_k / M_V  \ll 1$ such that the loop functions
$f,g,,h,j$ approach constant values \cite{Hati:2019ufv}.

We can express the contribution of the LQ to the lepton anomalous moment as
\begin{equation}
  \label{amu}
  \Delta a_\ell = - \frac{N_C}{16 \pi^2} \biggr [
     4 {\cal R}e ( X^{LR*}_{3 \ell} X^{RL}_{3 \ell} ) \frac{m_b m_\ell}{M_V^2} (Q_V + Q_{b^c} )
    + 2 (  |X^{LR}_{3\ell} |^2 + |X^{RL}_{3\ell} |^2 ) \frac{m_{\ell}^2 }{M_V^2}
    \left( \frac{5}{6} Q_V + \frac{2}{3} Q_{b^c} \right ) \biggr ]
\end{equation}
where we assume that the contribution from $b$ quark dominates over the $s$ and $d$ quarks. 
The partial width for the radiative decay of, such as $\mu \to e \gamma$, is given by
\begin{equation}
  \Gamma (\mu \to e \gamma) = \frac{ \alpha_{\rm em}}{4} m_{\mu}^5
  \left ( |A_L^{\mu e}  |^2 + |A_R^{\mu e} |^2 \right )
\end{equation}

\subsection{ $R_{K,K^*}$}

The effective Lagrangian for a generic exclusive decay of $b \to s \ell^- \ell^+$ ,
with $\ell  = e,\,\mu,\, \tau$ can be written as
\begin{equation}
  {\cal L}_{bs\ell\ell}  \supset  \frac{4 G_F}{\sqrt{2}}\, V_{tb} V^*_{ts} \, \frac{e^2}{16 \pi^2}
  \, \sum_i  \, \left [ C_i {\cal O}_i + C'_i {\cal O}'_i  \right ]  + h.c. \,,
\end{equation}
where
\begin{eqnarray}
&&  {\cal O}_9= ( \bar s \gamma_\mu P_L b )\, ( \bar \ell \gamma^\mu \ell )\,,  \qquad
  {\cal O}'_9= ( \bar s \gamma_\mu P_R b )\, ( \bar \ell \gamma^\mu \ell )\,,  \nonumber \\
  &&{\cal O}_{10}= ( \bar s \gamma_\mu P_L b )\, ( \bar \ell \gamma^\mu \gamma^5\ell )\,, \qquad
    {\cal O}'_{10}= ( \bar s \gamma_\mu P_R b )\, ( \bar \ell \gamma^\mu \gamma^5\ell )
      \,,  \nonumber \\
&&  {\cal O}_{S}= ( \bar s \ P_R b )\,  ( \bar \ell  \ell ) \,, \qquad
        {\cal O}'_{S}=  ( \bar s  P_L b )\, ( \bar \ell \ell )       \,,  \nonumber \\
&& 
  {\cal O}_{P}= ( \bar s  P_R b )\, ( \bar \ell \gamma_5  \ell )\,,  \qquad
  {\cal O}'_{P}= ( \bar s  P_L b )\, ( \bar \ell  \gamma_5\ell )      \,,  \nonumber
  \end{eqnarray}
Since the term responsible for the left-handed down-type quark $V_2$ interaction  is
$- X_{ij}^{LR} \, \overline{d_L^{c,i}} \,\gamma^\mu  e_R^j \, V^{+4/3}_\mu $,
which involves only the right-handed charged lepton $e_R^j$, it will give rise to the
relation $C_9 = C_{10}$ between $C_9$ and $C_{10}$.  Therefore, we focus the
interactions with the electron instead of muon, as the best fit results preferred
$C_9 = - C_{10}$ for muon but $C_9 = + C_{10}$ for electron.
The contributions to the Wilson coefficients from $V_2$ focusing on the interactions
with electron are given by
\begin{eqnarray} 
C^{bsee}_9 = + C_{10} ^{bsee}  &=& - \frac{4\pi^2}{e^2} \, \frac{v^2}{M^2_{V_2}} \,
\frac{ X^{LR}_{31} X^{LR*}_{21} }{V_{ts}^* V_{tb}} \;, \nonumber \\
C^{' bsee }_9 = - C_{10} ^{' bsee}  &=& - \frac{4\pi^2}{e^2} \, \frac{v^2}{M^2_{V_2}} \,
\frac{ X^{RL}_{31} X^{RL*}_{21} }{V_{ts}^* V_{tb}} \;, \nonumber \\
C^{ bsee }_S = - C_{P} ^{ bsee}  &=&  \frac{4\pi^2}{e^2} \, \frac{2 v^2}{M^2_{V_2}}  \,
\frac{ X^{RL}_{31} X^{LR*}_{21} }{V_{ts}^* V_{tb}} \;, \nonumber \\
C^{' bsee }_S = + C_{P} ^{' bsee}  &=&
 \frac{4\pi^2}{e^2} \, \frac{2 v^2}{M^2_{V_2}} \,
\frac{ X^{LR}_{31} X^{RL*}_{21} }{V_{ts}^* V_{tb}} \;, \nonumber 
\end{eqnarray}
Strictly speaking, the above Wilson coefficients are calculated at the electroweak scale by
integrating out the heavy degrees of freedom such as the LQ. One has to evolve them down
to the $m_b$ scale.  However, giving that the evolution effect is very small because of 
small mixings between the operators and 
${\cal O}_2 \equiv  (\bar s \gamma_\mu ( 1- \gamma_5) c)
                                      (\bar c \gamma^\mu (1- \gamma_5) b)$,
\footnote{We specially thank Wolfgang Altmannshofer for explaining this point.}
we directly employ the above expressions in our analysis, similarly for the Wilson
coefficient $C_{S_R}$.

        
\subsection{$ R_{D, D^*}$ }

The effective Lagrangian for a generic exclusive decay of $b \to c  \tau ^-  \bar \nu_\tau$
can be written as
\begin{eqnarray}
  {\cal L}_{bc\ell\nu} &=&  - 2 \sqrt{2} G_F V_{cb} \, \biggr [
    (1 + C_{V_L} )( \overline{c_L} \gamma^\mu b_L ) \, ( \overline{\tau} \gamma^\mu
      \nu_{\tau_L} )
    + C_{V_R} ( \overline{c_R} \gamma^\mu b_R ) \, ( \overline{\tau} \gamma^\mu \nu_{\tau_L} )
    \nonumber \\
    &+&
    C_{S_R} ( \overline{c_L} b_R ) \, ( \overline{\tau}  \nu_{\tau_L } )
    + C_{S_L} ( \overline{c_R} b_L ) \, ( \overline{\tau}  \nu_{\tau_L}  )
    + C_{T} ( \overline{c_R} \sigma^{\mu\nu} b_L ) \, ( \overline{\tau} \sigma_{\mu\nu}
    \nu_{\tau_L}  ) \; \biggr ]
\end{eqnarray}
The isospin $- 1/2$ component of the $V_2$ contributes to the process $ b \to c \tau \nu_{\tau}$
via the operator $ ( \overline{c_L} b_R ) \, ( \overline{\tau}  \nu_{\tau_L } )$. and thus
modifies the coefficient $C_{S_R}$:
\begin{equation}
\label{eq:ClSR}
  C_{S_R} = - \frac{1}{2 \sqrt{2} G_F} \, \frac{1}{V_{cb} } \,
   \sum_k \left( V_{k2} \, \frac{2}{M_{V}^2 }   \, X^{RL}_{33} X^{LR*}_{k3} \right  )
\end{equation}

\subsection{$B_s \to \ell^+\ell^-$}

Including the new physics contributions, 
the general expression of the $B_s \to \ell^+\ell^-$ in terms of Wilson coefficients is
~\cite{Kosnik:2012dj}
\begin{eqnarray}
{\rm Br}(B_s \to \ell^+ \ell^-)&=& \tau_{B_s} f^2_{B_s}m^3_{B_s} 
\frac{G^2_F |V_{tb}V^*_{ts}|^2 e^4}{(4 \pi)^5} \sqrt{1-4m^2_\ell/m^2_{B_s}} \nonumber \\
& \times & \left[
  \frac{m^2_{B_s}}{m^2_b} \left\vert C_S-C'_S \right\vert^2 \left( 1-\frac{4m^2_\ell}
       {m^2_{B_s}} \right) 
+\left\vert  \frac{m_{B_s}}{m_b}\left( C_P-C'_P \right)
+\frac{2m_\ell}{m_{B_s}}\left( C^{\rm SM}_{10}+C_{10}-C'_{10} \right)  \right\vert^2
\right]\,, \nonumber \\
\end{eqnarray}
where $C^{\rm SM}_{10}=-4.1$ stems from the SM contribution, 
$\tau_{B_s}=1.52\times 10^{-12}~{\rm s}$ and $f_{B_s}=228.4~{\rm MeV}$ 
are the lifetime and decay constant of the $B$-meson, respectively.
The current measurements are~\cite{Altmannshofer:2021qrr}
\begin{eqnarray}
&& {\rm Br}(B_s \to \mu^+ \mu^-)=(3.09^{+0.48}_{-0.44})\times 10^{-9}\,, \\
&& {\rm Br}(B_s \to e^+ e^-)< 9.4 \times 10^{-9}
~~\text{at $90\%$ C.L from PDG},
\end{eqnarray}
and consistent with the SM.

\subsection{$W$ boson mass}

The $W$-boson mass due to new physics
expressed in terms of the $\Delta S, \Delta T, \Delta U$ is given by
\begin{equation}
  M_W^2 = M^2_{W,SM} + \frac{\alpha c_W^2 M_Z^2 }{ c_W^2 - s_W^2} \,
  \left[ - \frac{\Delta S}{2} + c_W^2 \Delta T + \frac{c_W^2 - s_W^2 }{4 s_W^2} \Delta U
    \right ]
\end{equation}
where $c_W, s_W$ are, respectively, the cosine and the sine of the Weinberg angle.
$\alpha$ is the fine-structure constant, and  $M_Z$ is the $Z$-boson mass.  In the SM,
$M_{W,SM} = 80.361 \pm 0.006$ GeV \cite{ParticleDataGroup:2016lqr},
while the most recent measurement by CDF 
is $M_W = 80.4335 \pm 0.0094 $ GeV \cite{CDF:2022hxs}.
Such a large deviation can be most easily
accommodated by a positive $\Delta T$.  It is well know that
an additional weak doublet with a mass splitting gives a contribution to $\Delta T$ as
\footnote{This expression is based on the formulas listed in the Appendix D of the Stefan Pokorski's book~\cite{Pokorski:1987ed}.}
\begin{eqnarray}
  \Delta T &=& \frac{C}{4 \pi M_W^2 s_W^2} \, F(m_1, m_2) \;,
\end{eqnarray}
where $C=1 (3)$ for color singlet (triplet) and 
\[
F(m_1, m_2 ) = m_1^2 + m_2^2 - \frac{2 m_1^2 m_2^2}{m_1^2 - m_2^2} \log
\left ( \frac{m_1^2 }{m_2^2} \right ) \;.
\]
Note that $ \Delta T$ is proportional to the mass splitting $\Delta m = |m_1 - m_2|$
between the upper and the lower component of the doublet, and
thus always positive.
With this
$\Delta T$ the change in $M_W^2$ is given by
\begin{equation}
  \label{raise}
  \Delta M_W^2 = \frac{ \alpha c_W^4 M_Z^2 }{ c_W^2  - s_W^2} \Delta T
\end{equation}

\section{Parameter scans and Results}

\begin{table}[tb!]
  \centering
  \caption{Best fitted results of the relevant Wilson coefficients for 
  $R_{K,K^*}$~\cite{Altmannshofer:2021qrr} and
    $R_{D,D^*}$~\cite{Bhol:2021iow}, muon and electron  anomalous magnetic moments $\Delta a_\mu$~\cite{Muong-2:2021ojo}
    and $\Delta a_e$~\cite{Aoyama:2012wj,Aoyama:2019ryr,Hanneke:2008tm}. 
    Note that there is a controversy for $\Delta a_e$ due to two
    different experimental measurements.
    \label{table-data} }
  \medskip
  \begin{tabular}{ll}
    \hline
        \hline
    $C_9^{bsee} = C_{10}^{bsee} $ &  $-1.28\,^{+0.24}_{-0.23}$\\
    $C^\ell_{S_R}$                &  $0.027\,^{+0.025}_{-0.026}$ \\
    $\Delta a_\mu $                              & $(251 \pm 59 ) \times 10^{-11}$ \\
    $\Delta a^{\rm LKB}_e $                  & $(4.8 \pm 3.0 ) \times 10^{-13}$ \\
    $\Delta a^{\rm B}_e $                    & $(- 8.8 \pm 3.6 ) \times 10^{-13}$ \\
    \hline \hline
  \end{tabular}
\end{table}

The best fitted results on the Wilson coefficients were updated in early 2021
\cite{Altmannshofer:2021qrr}.  We list them in Table~\ref{table-data}.
The LQ couplings can involve different leptonic flavors in the same Feynman diagram,
and so it will also give rise to the lepton-flavor-violating processes, notably, the
radiative leptonic decays $\ell_i \to \ell_j \gamma$.  While these processes have not
been observed, the upper limits are quite restrictive:
\begin{eqnarray}
  B (\tau \to \mu \gamma) < 4.4 \times 10^{-8}\,, && 90\%\; {\rm C.L.} \nonumber \\
  B (\tau \to e \gamma)  < 3.3 \times 10^{-8}\,, && 90\%\; {\rm C.L.} \nonumber \\
  B (\mu \to e \gamma) < 4.2 \times 10^{-13}\,, && 90\%\; {\rm C.L.} \nonumber
\end{eqnarray}

\begin{figure}[t!]
\centering
\includegraphics[height=1.1in,angle=0]{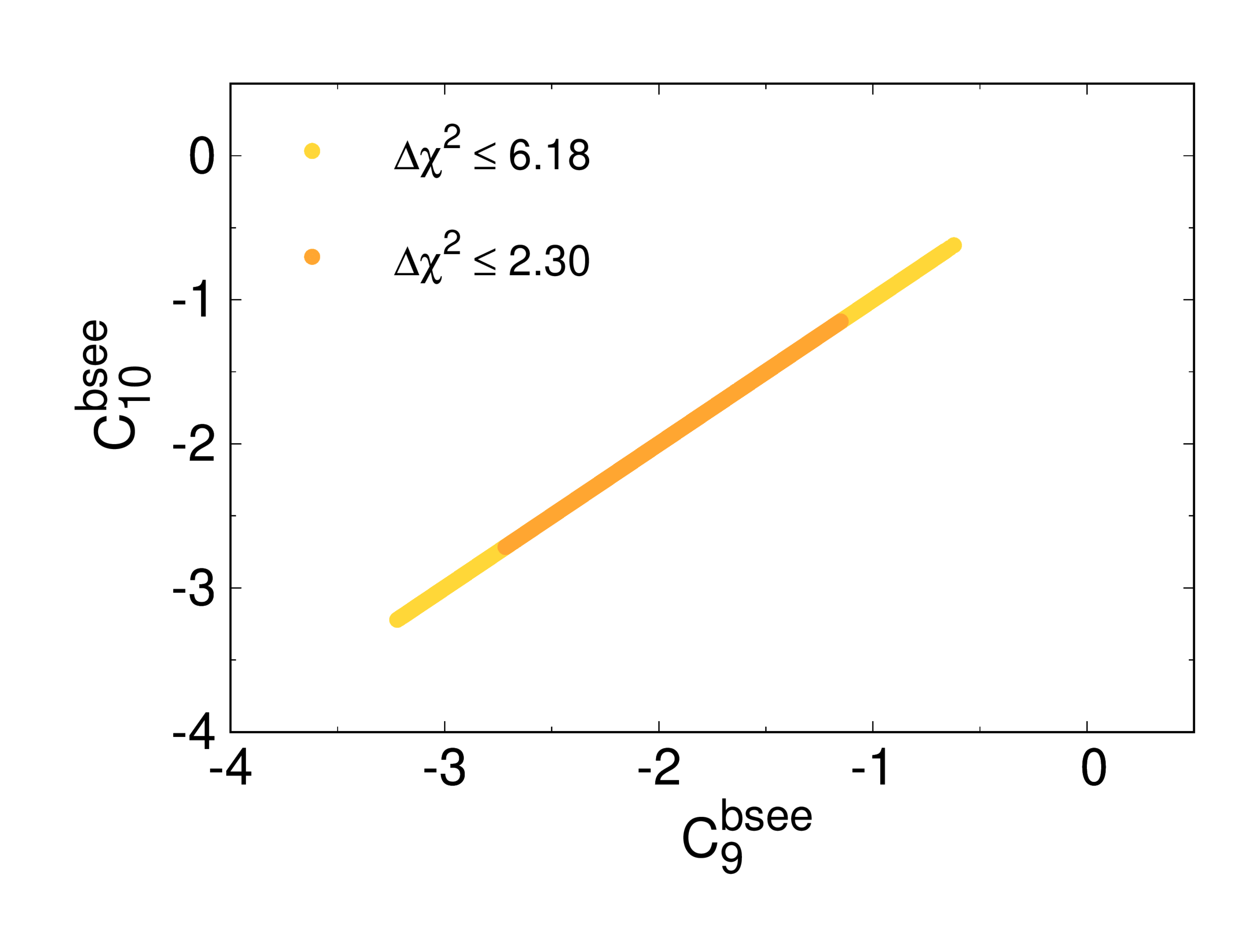}
\includegraphics[height=1.1in,angle=0]{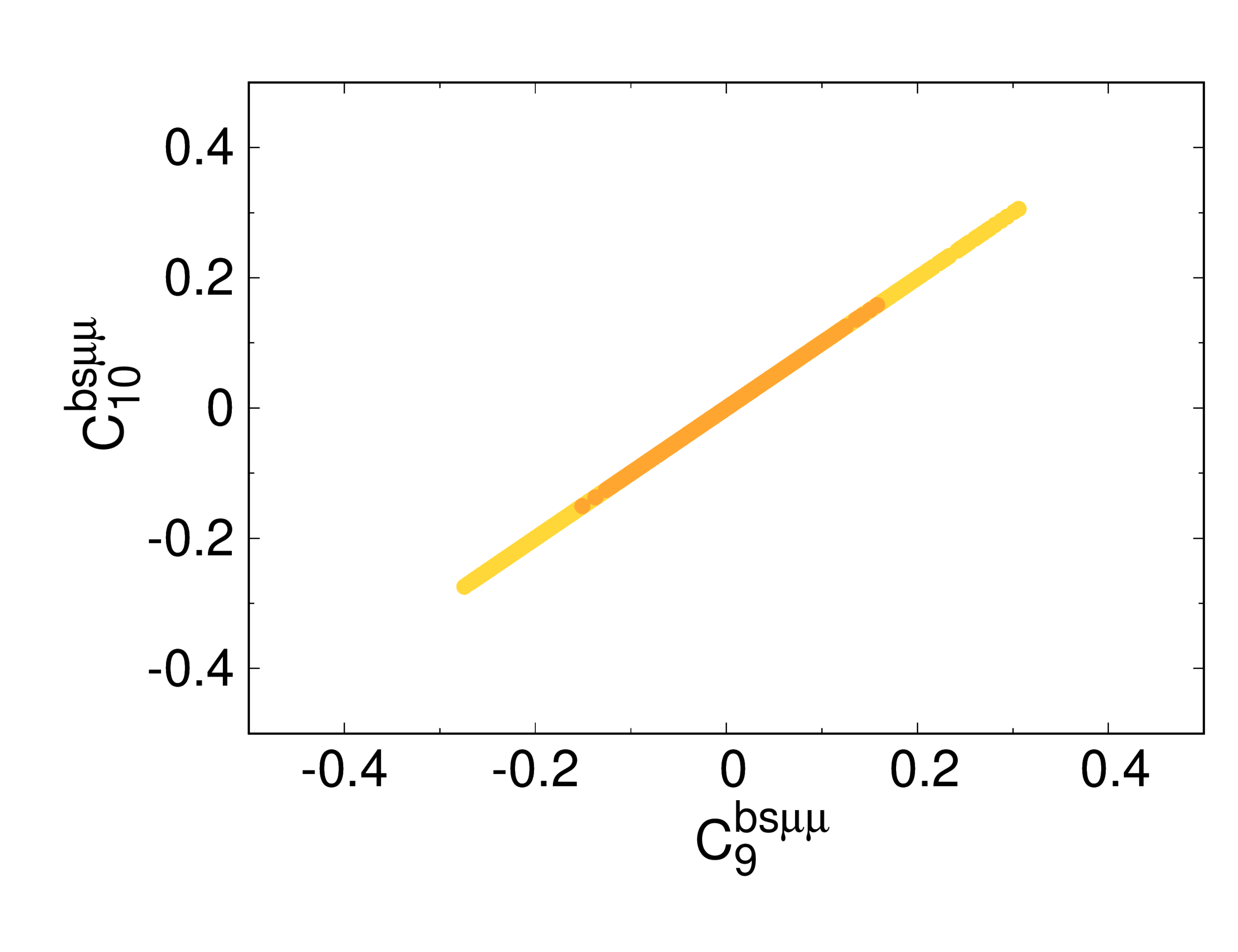}
\includegraphics[height=1.1in,angle=0]{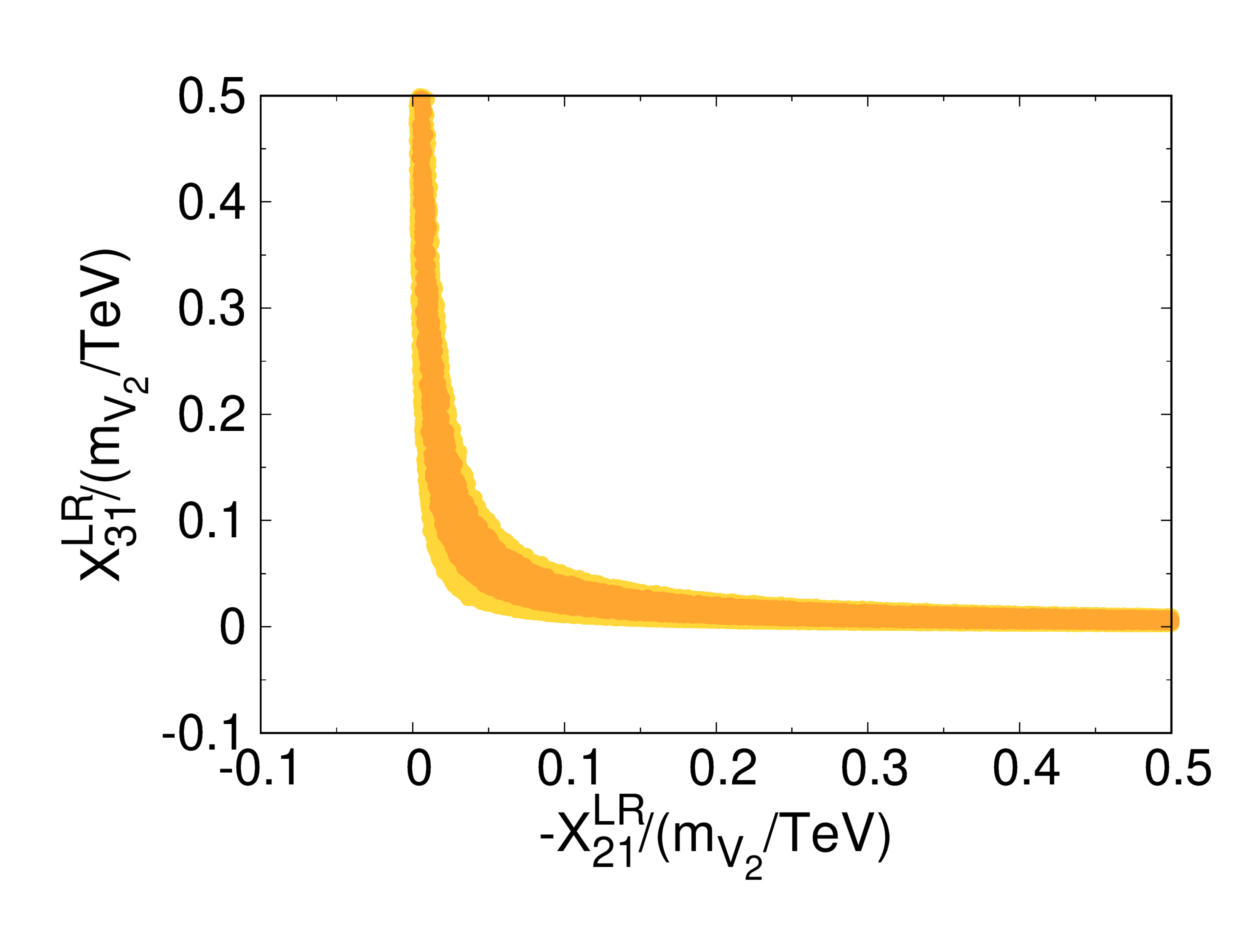}
\includegraphics[height=1.1in,angle=0]{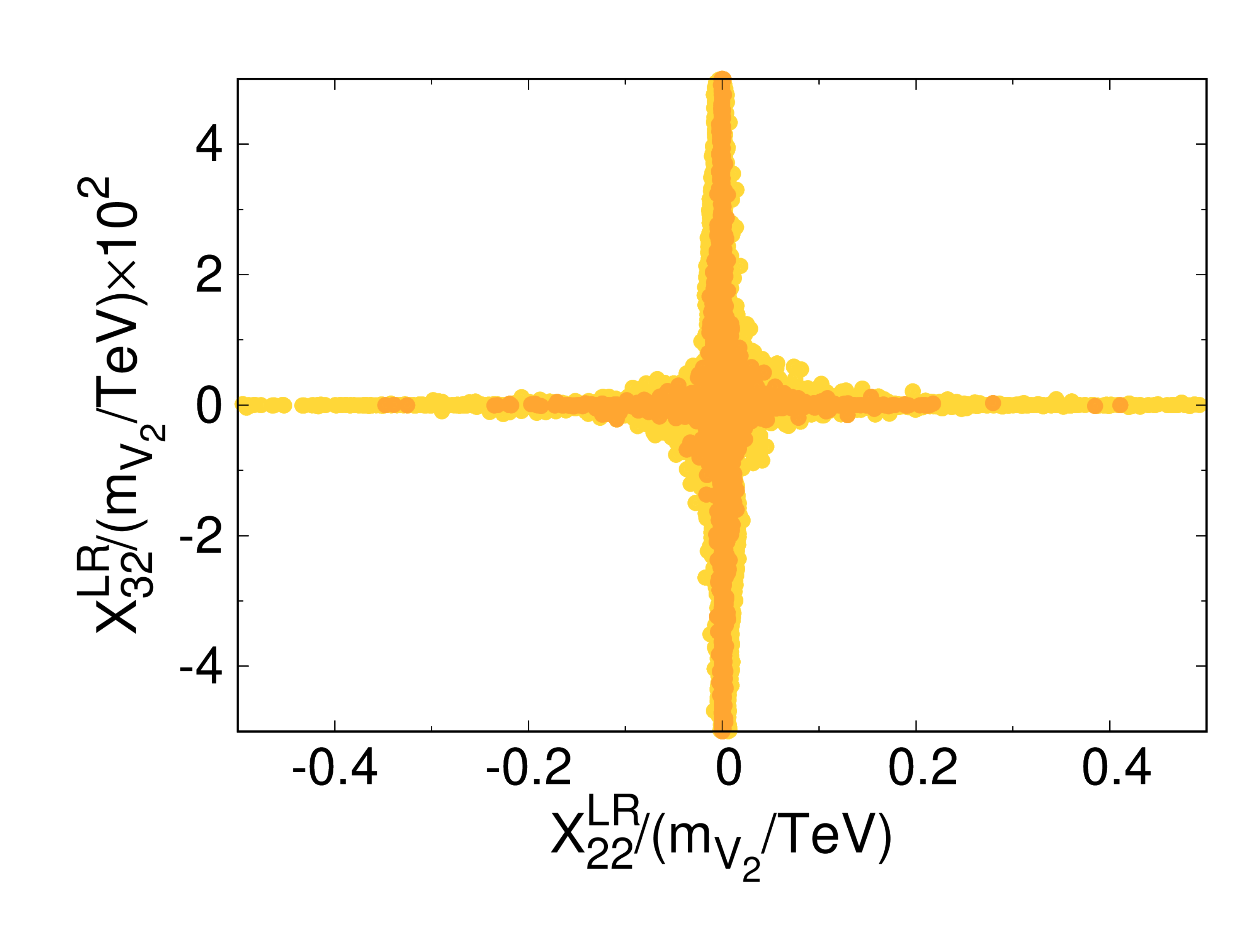}
\includegraphics[height=1.1in,angle=0]{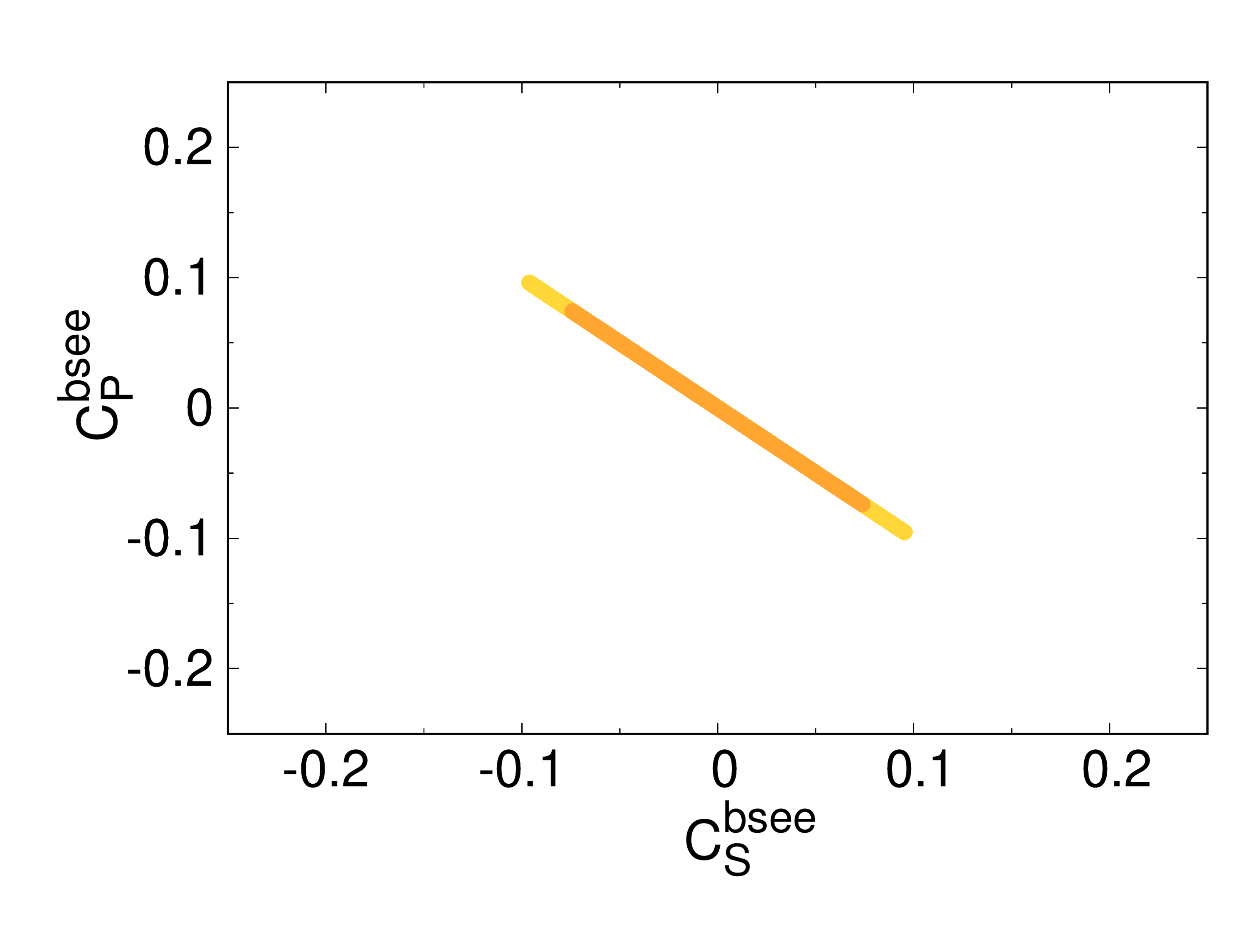}
\includegraphics[height=1.1in,angle=0]{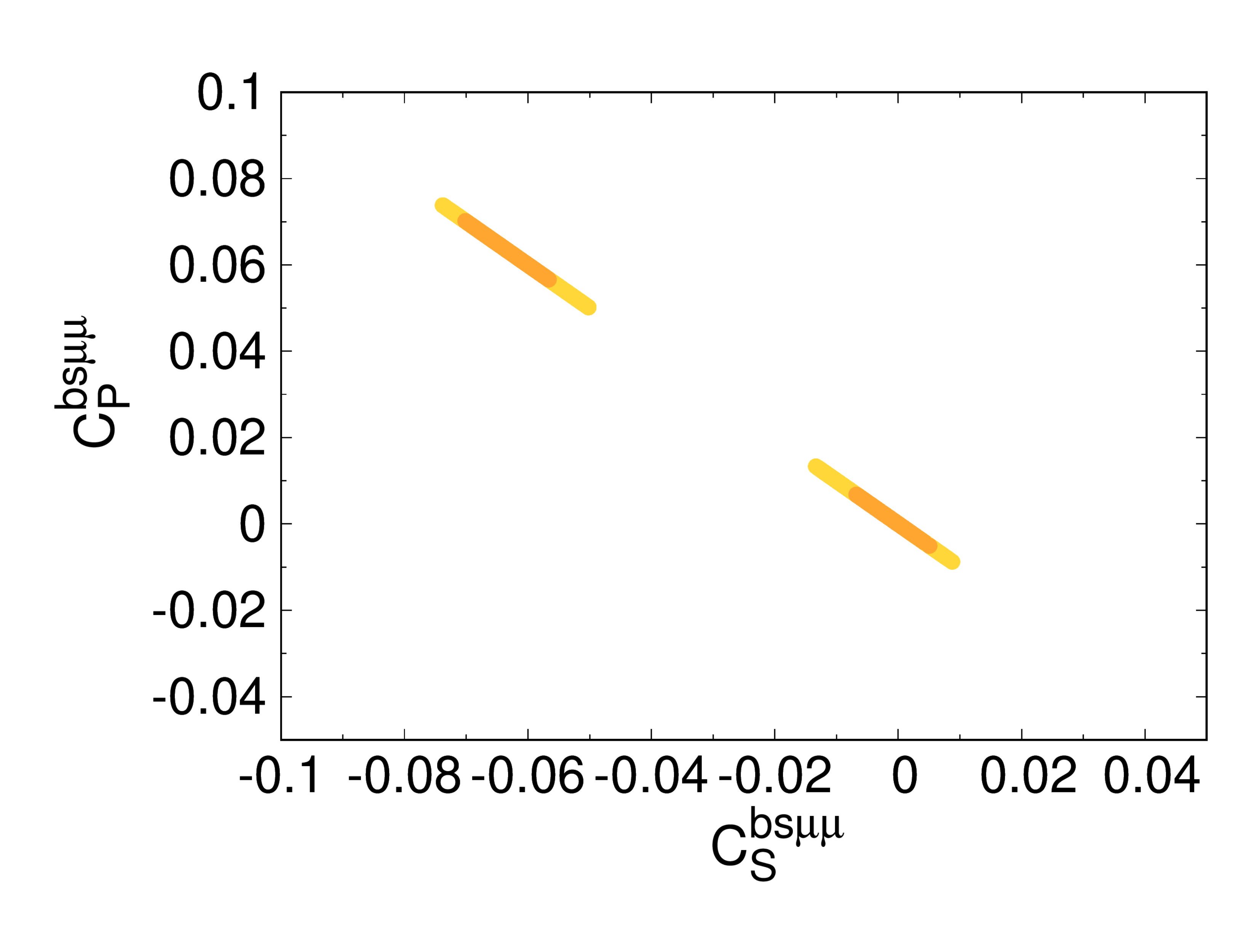}
\includegraphics[height=1.1in,angle=0]{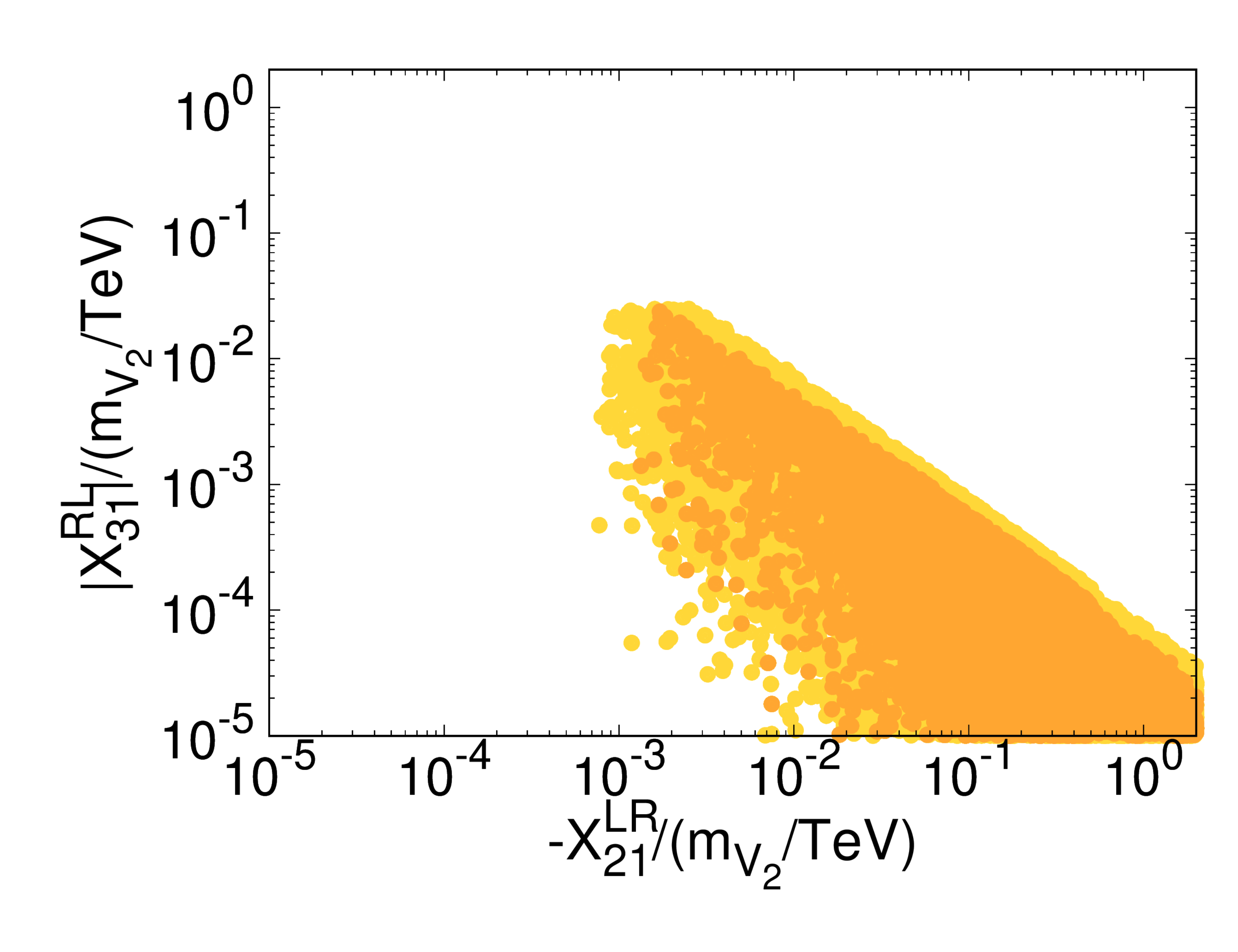}
\includegraphics[height=1.1in,angle=0]{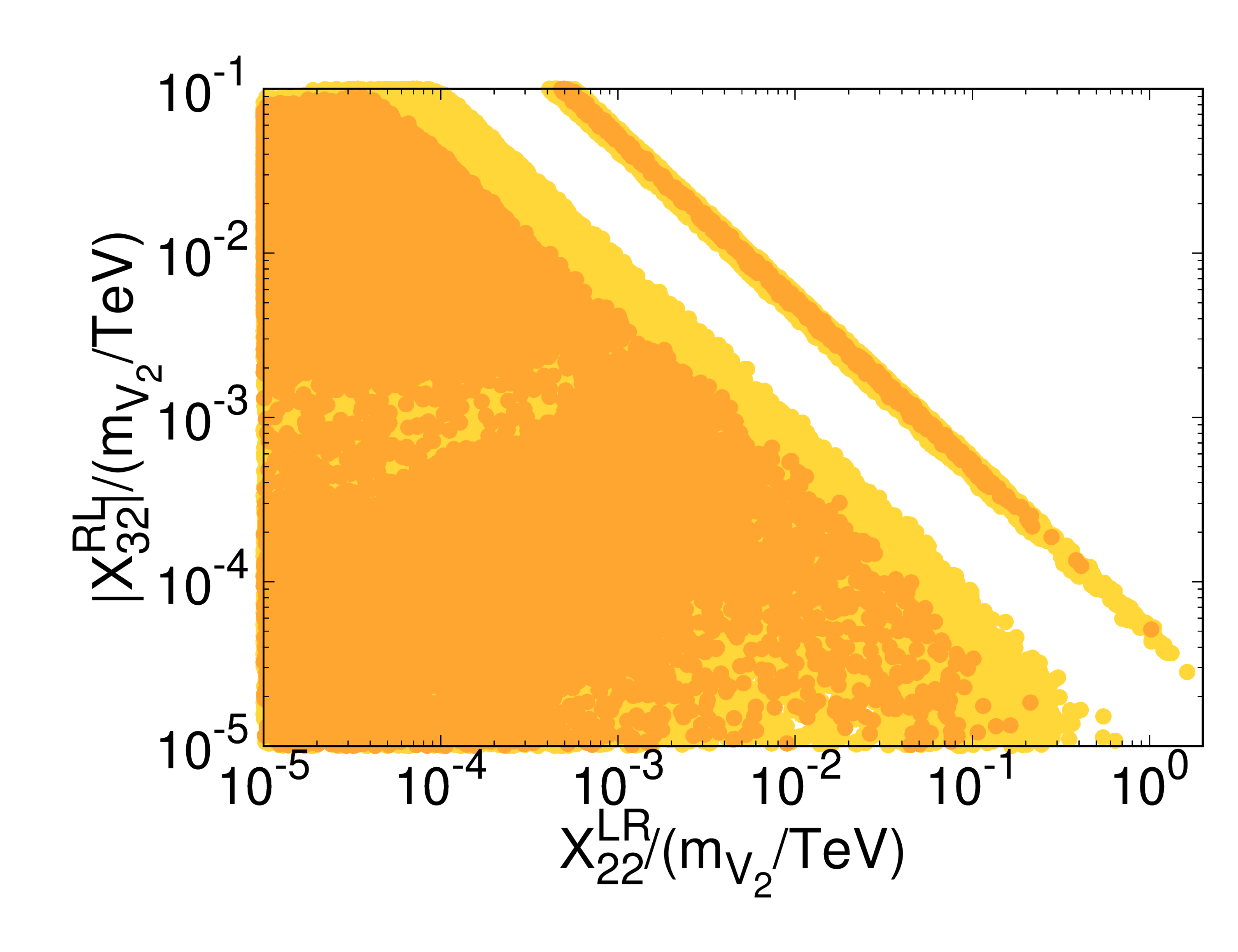}
\includegraphics[height=1.1in,angle=0]{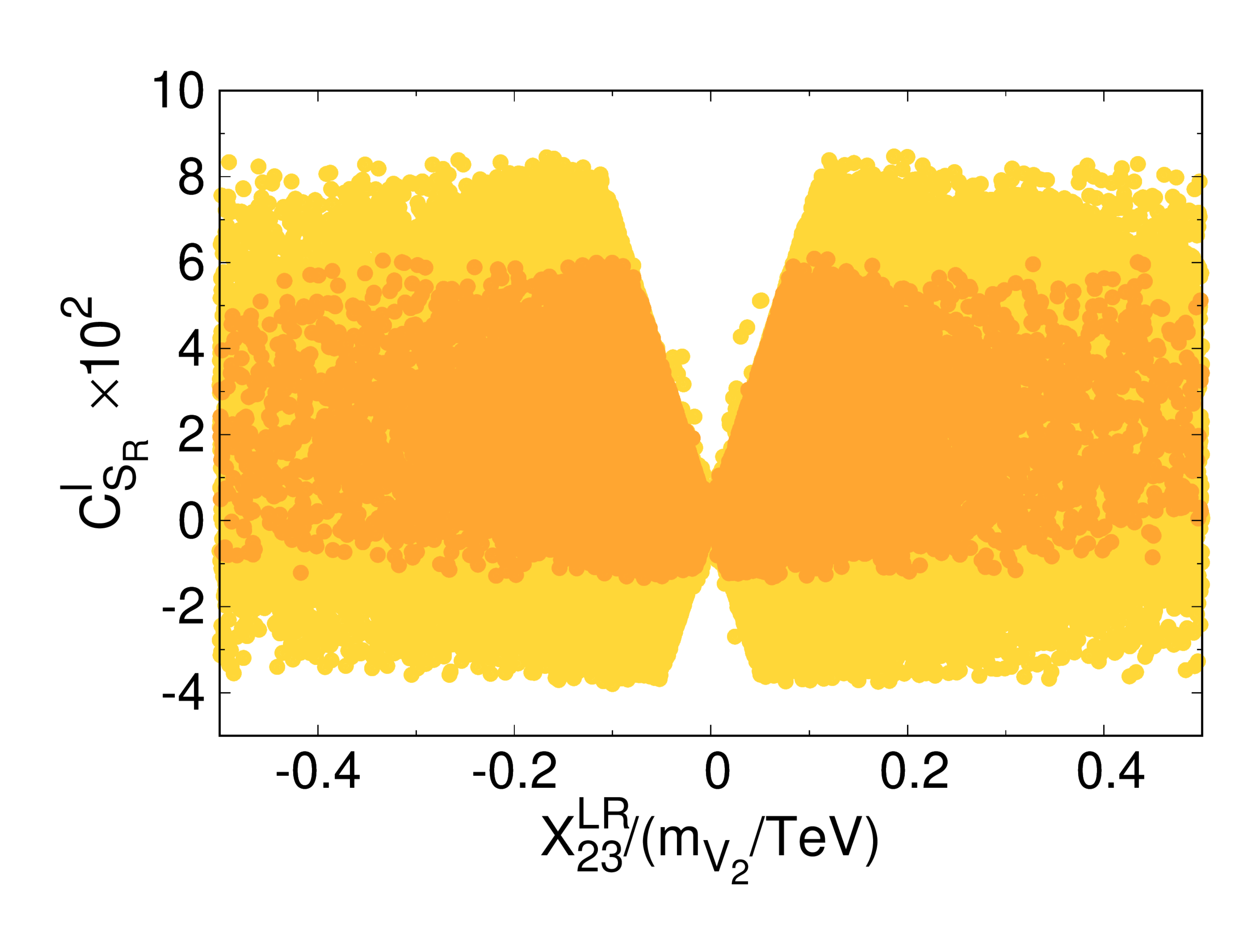}
\includegraphics[height=1.1in,angle=0]{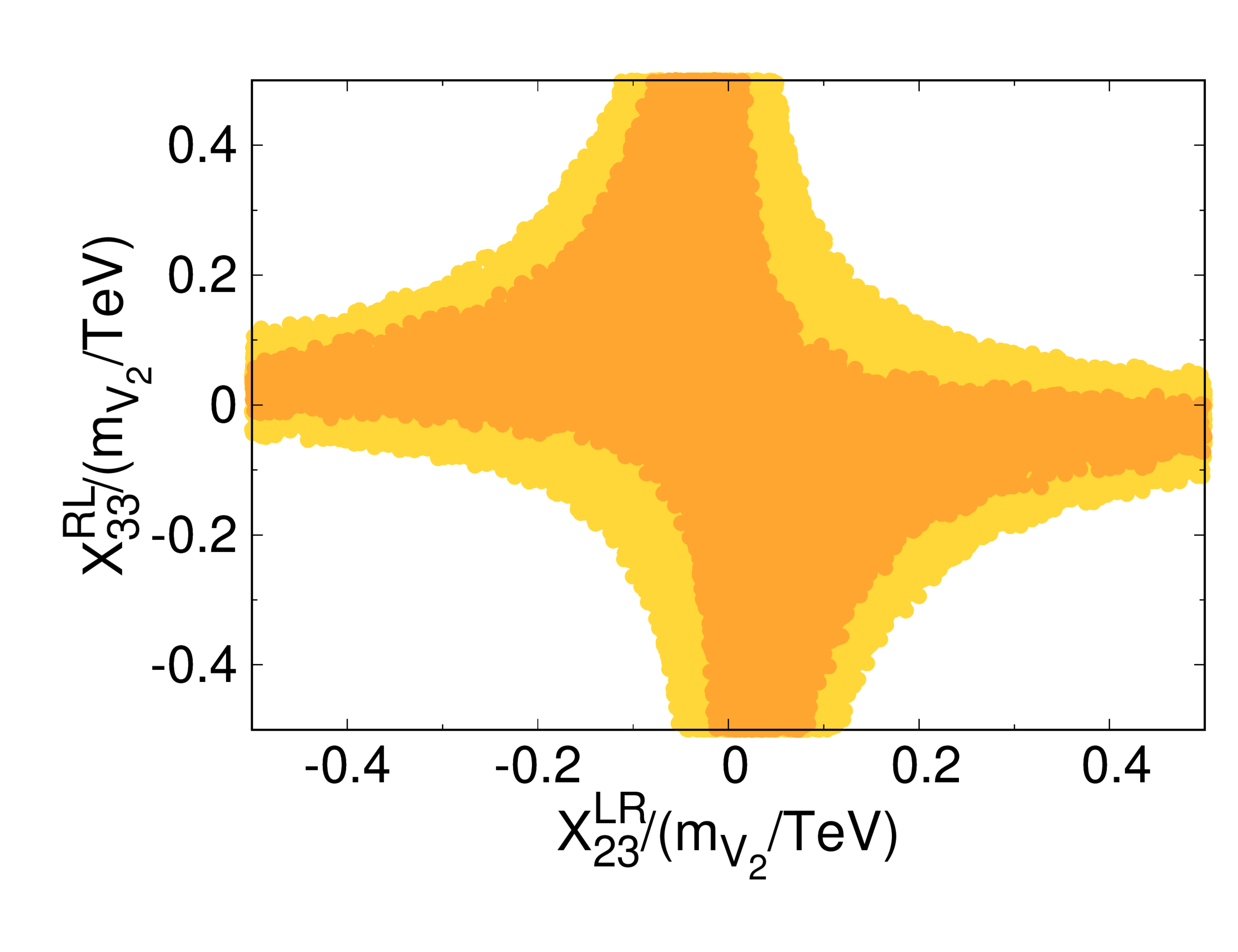}
\includegraphics[height=1.1in,angle=0]{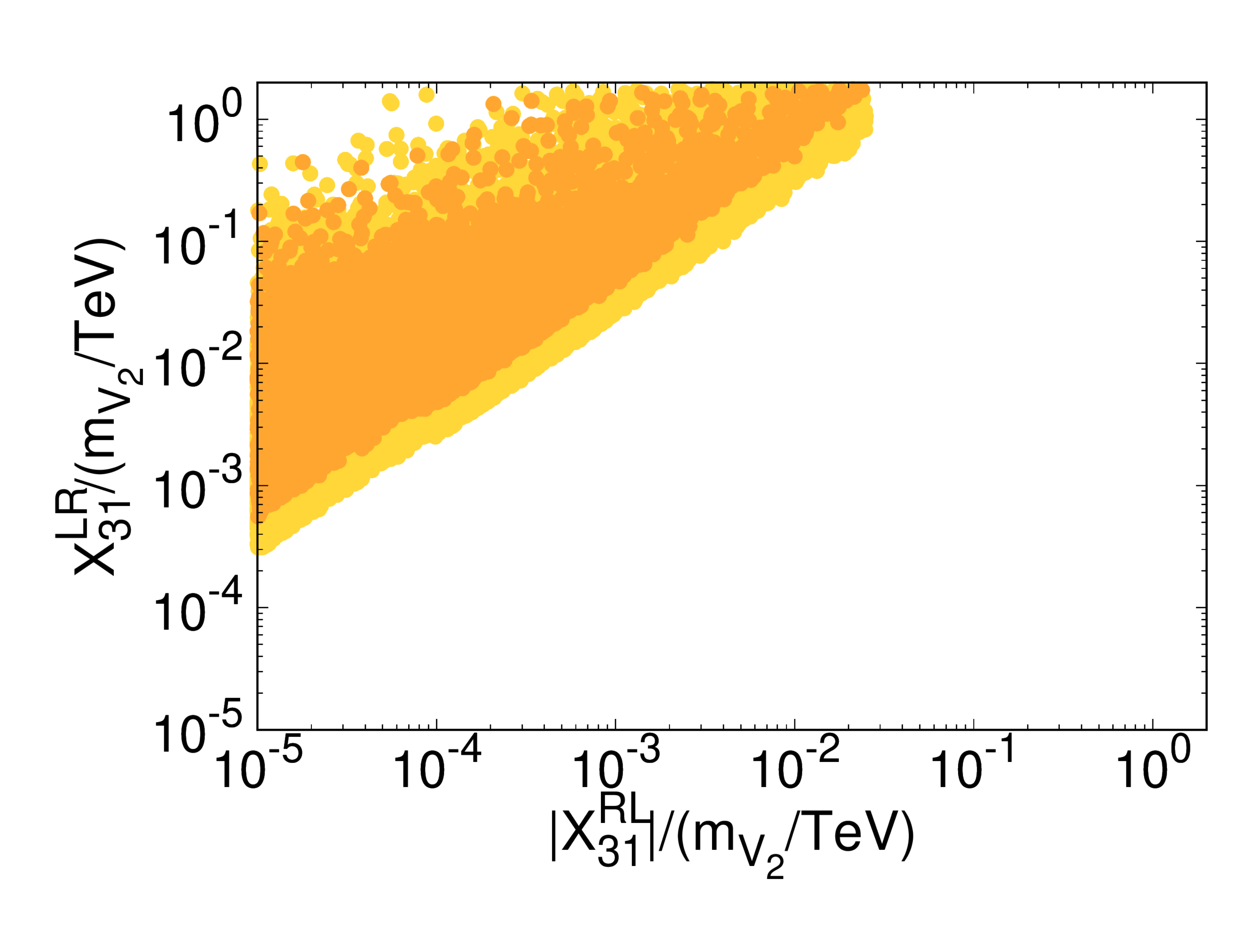}
\includegraphics[height=1.1in,angle=0]{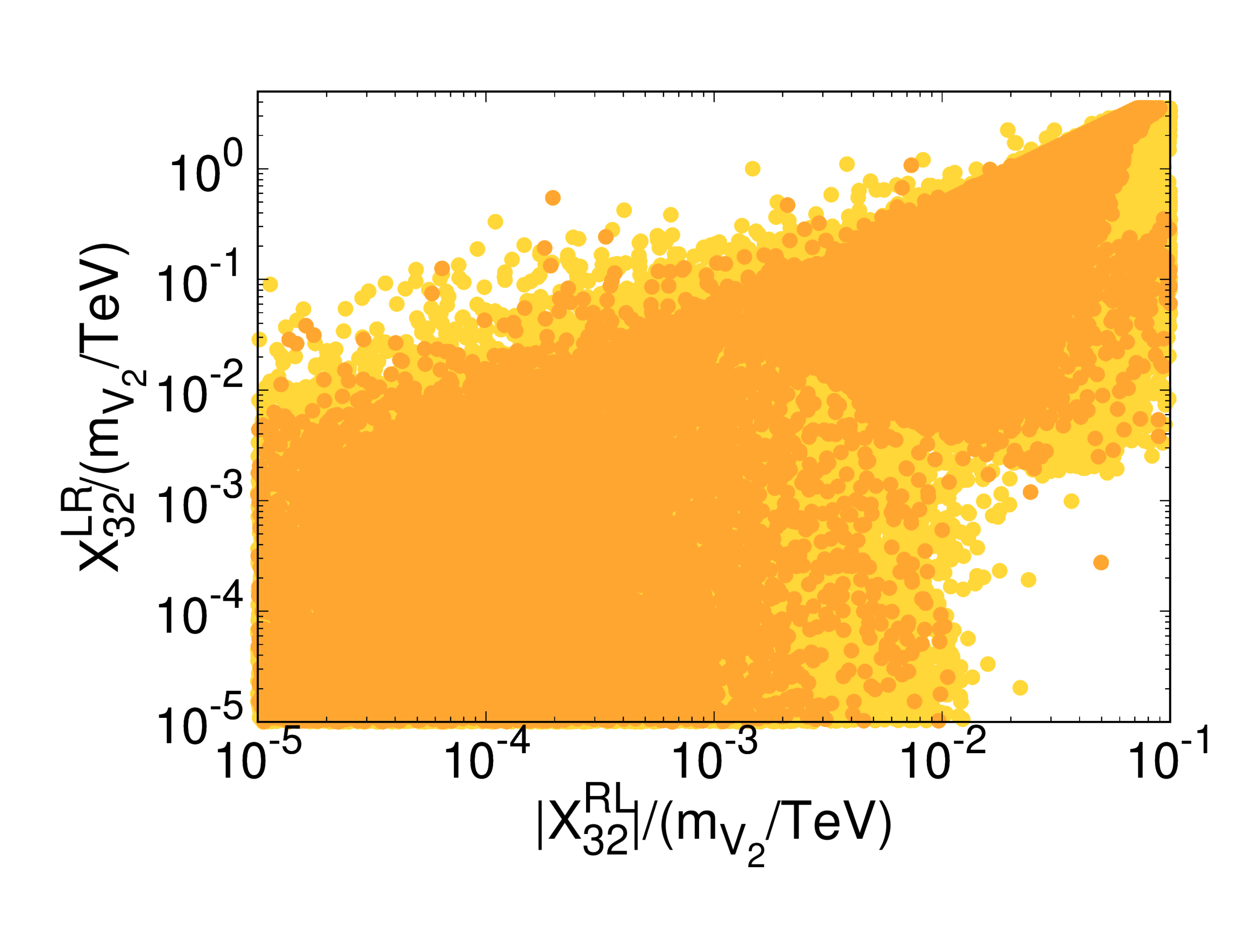}
\includegraphics[height=1.1in,angle=0]{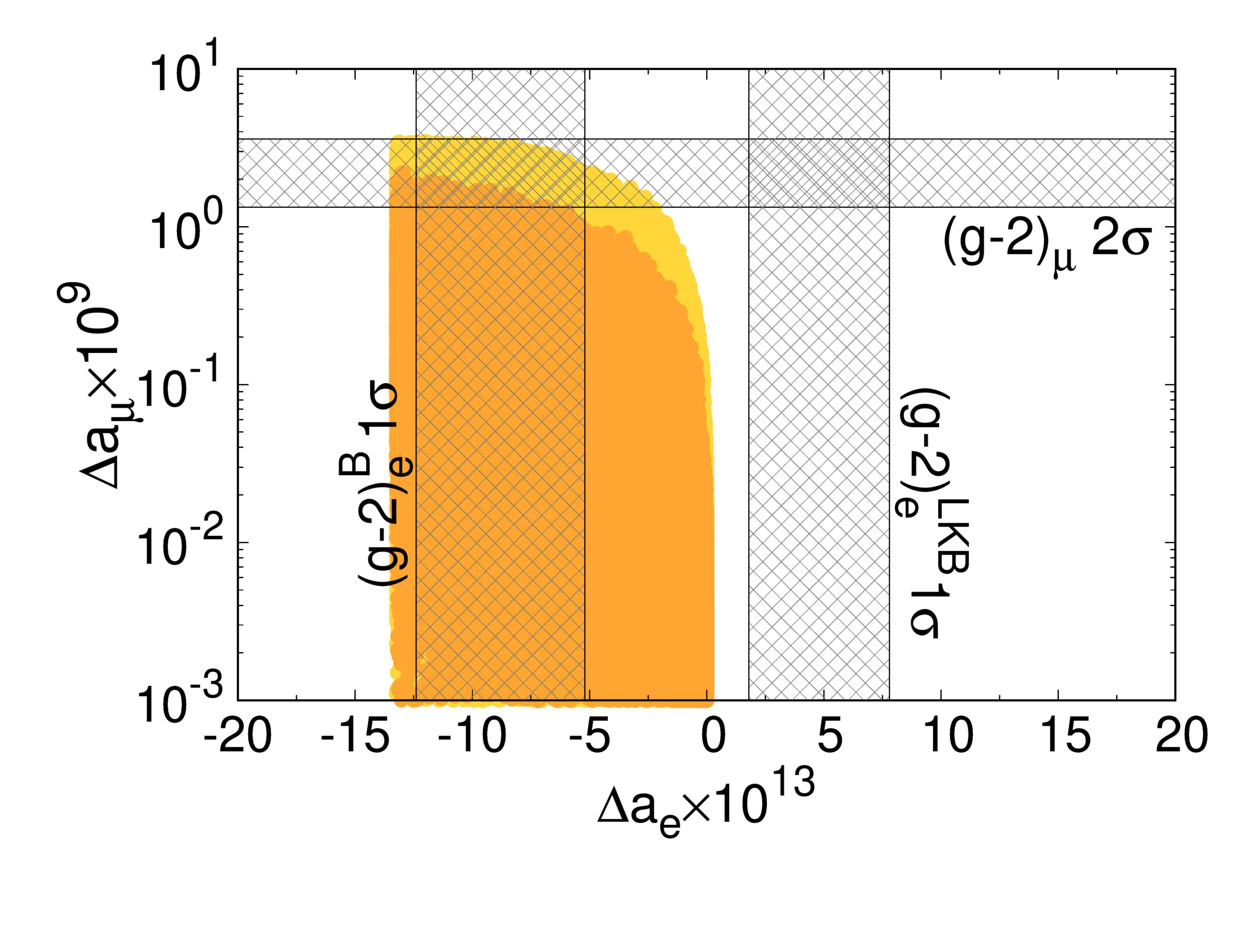}
\includegraphics[height=1.1in,angle=0]{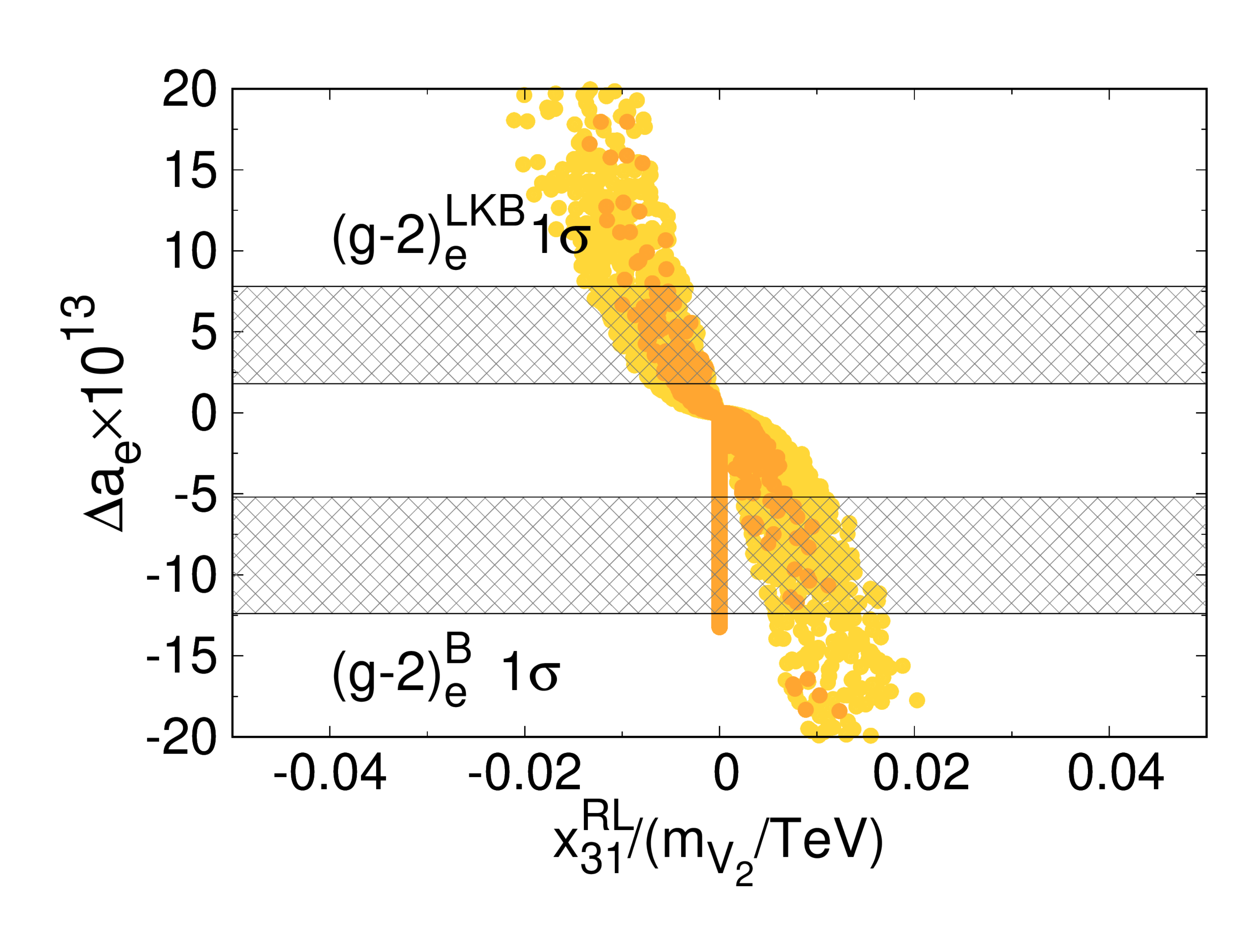}
\includegraphics[height=1.1in,angle=0]{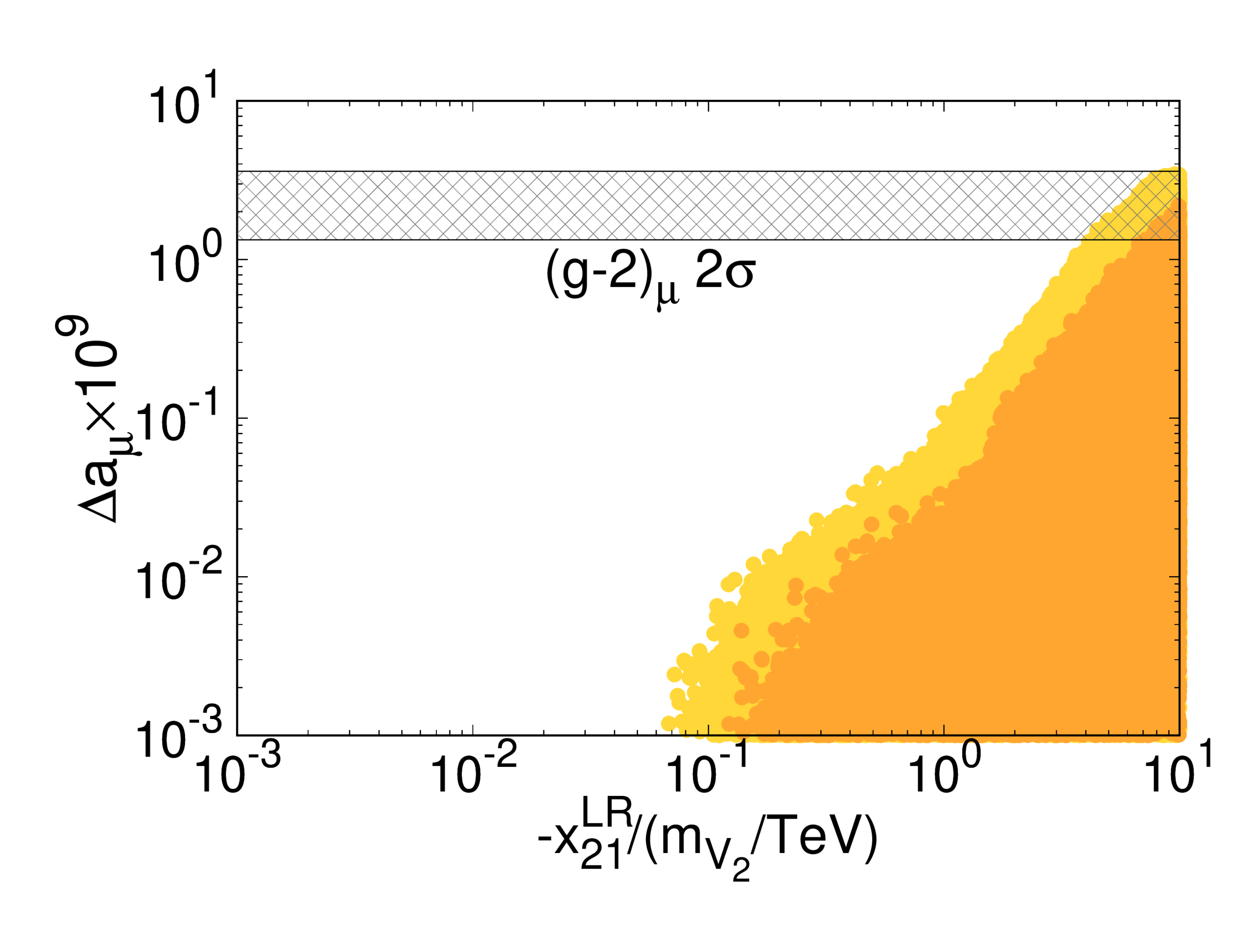}
\includegraphics[height=1.1in,angle=0]{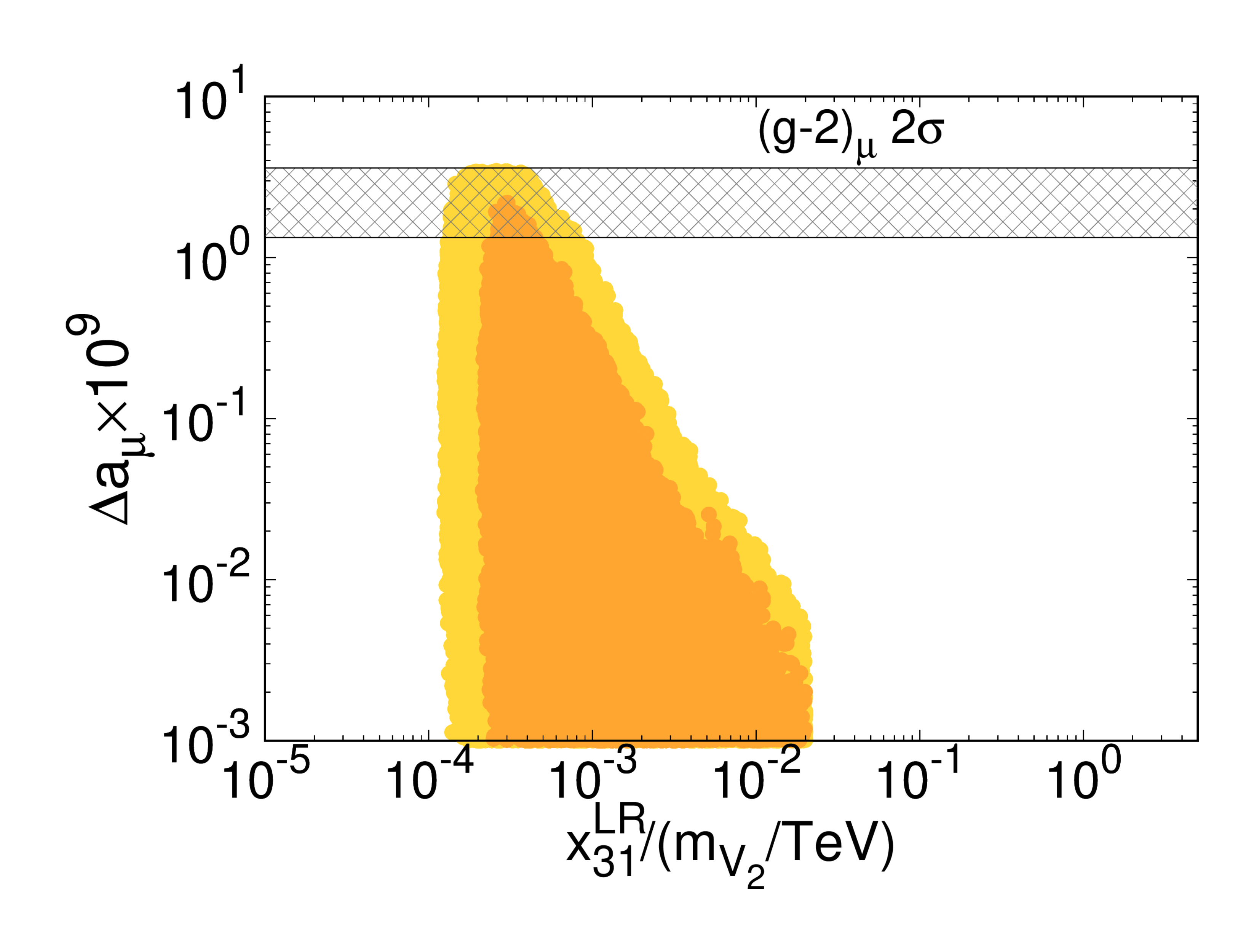}
\includegraphics[height=1.1in,angle=0]{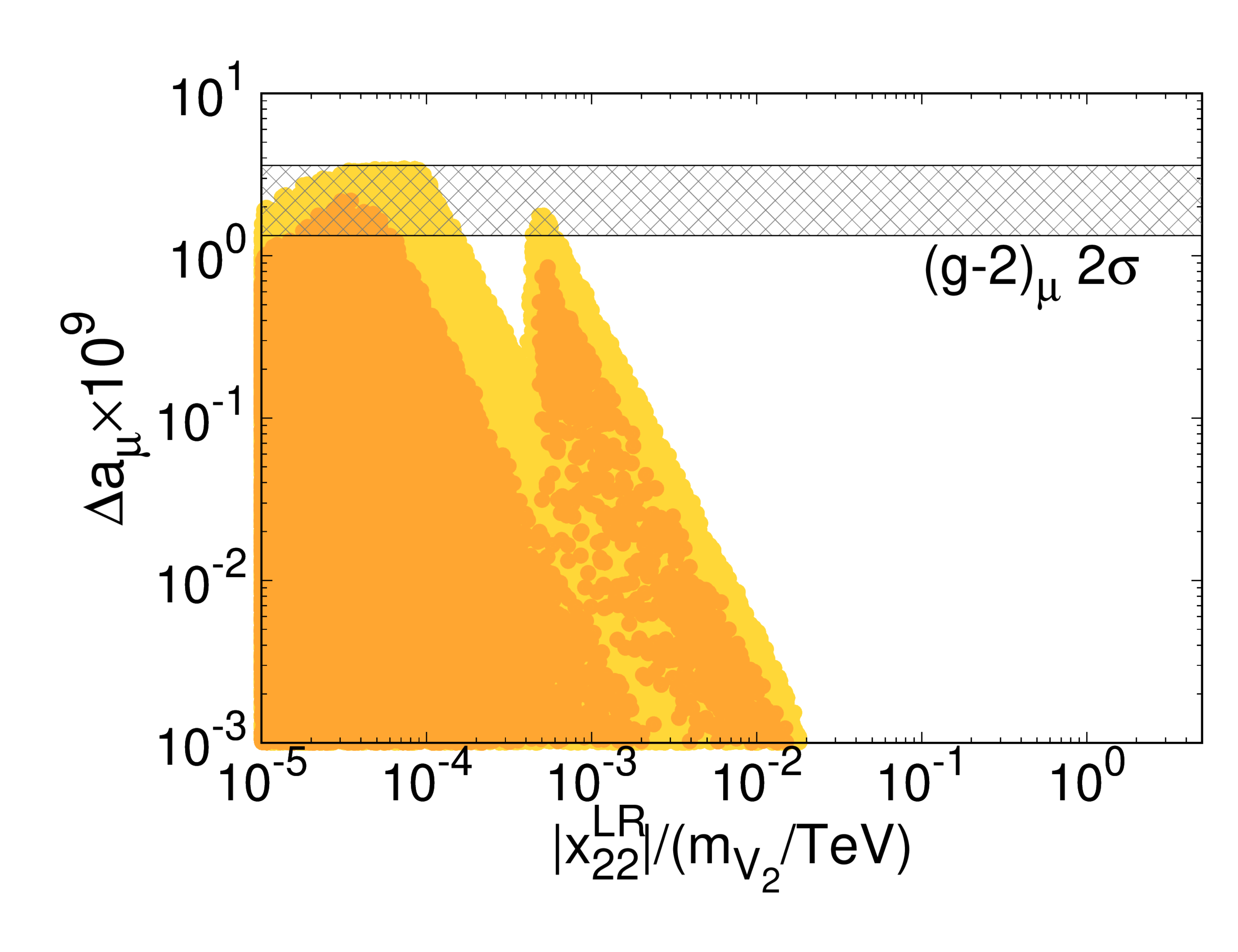}
\includegraphics[height=1.1in,angle=0]{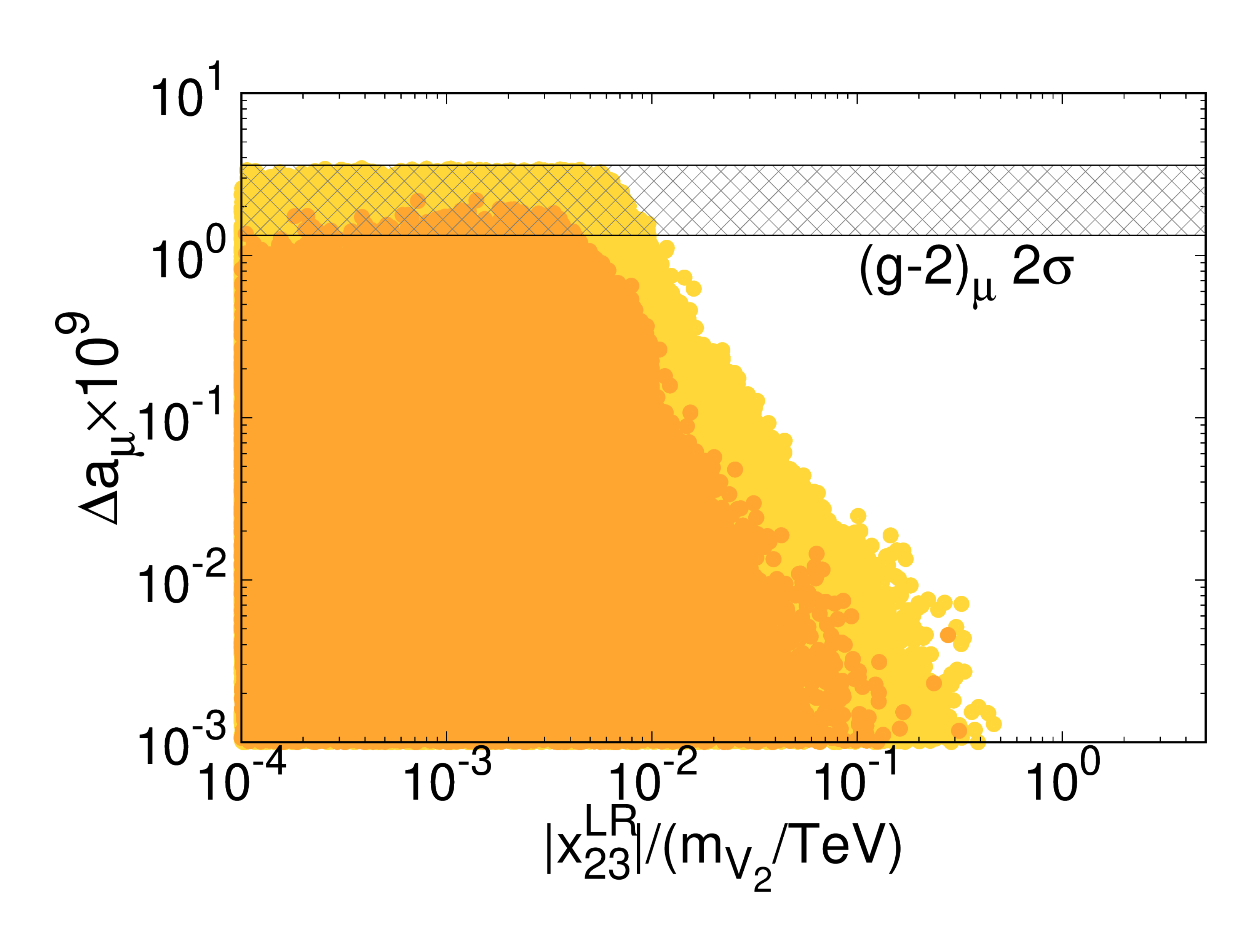}
\includegraphics[height=1.1in,angle=0]{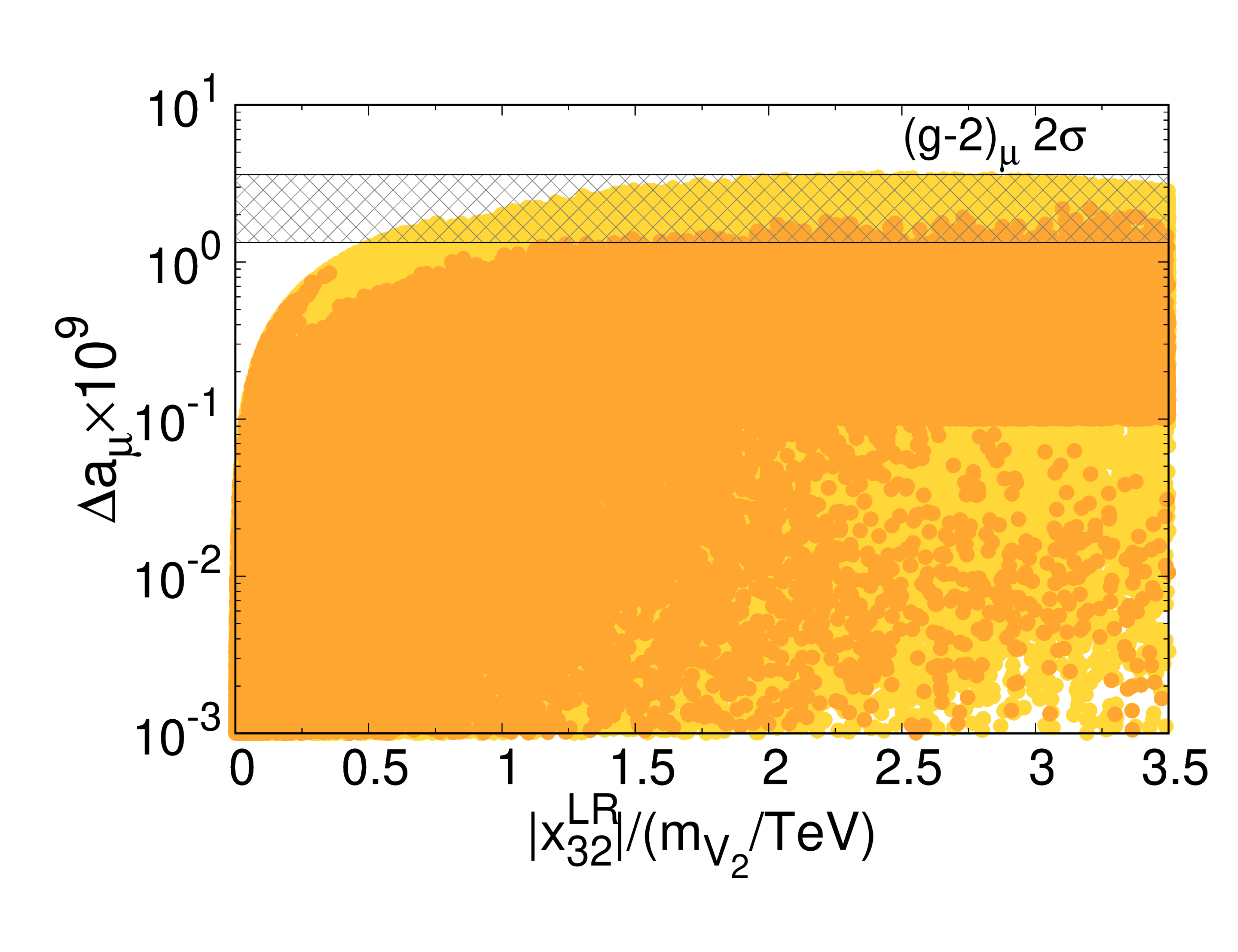}
\includegraphics[height=1.1in,angle=0]{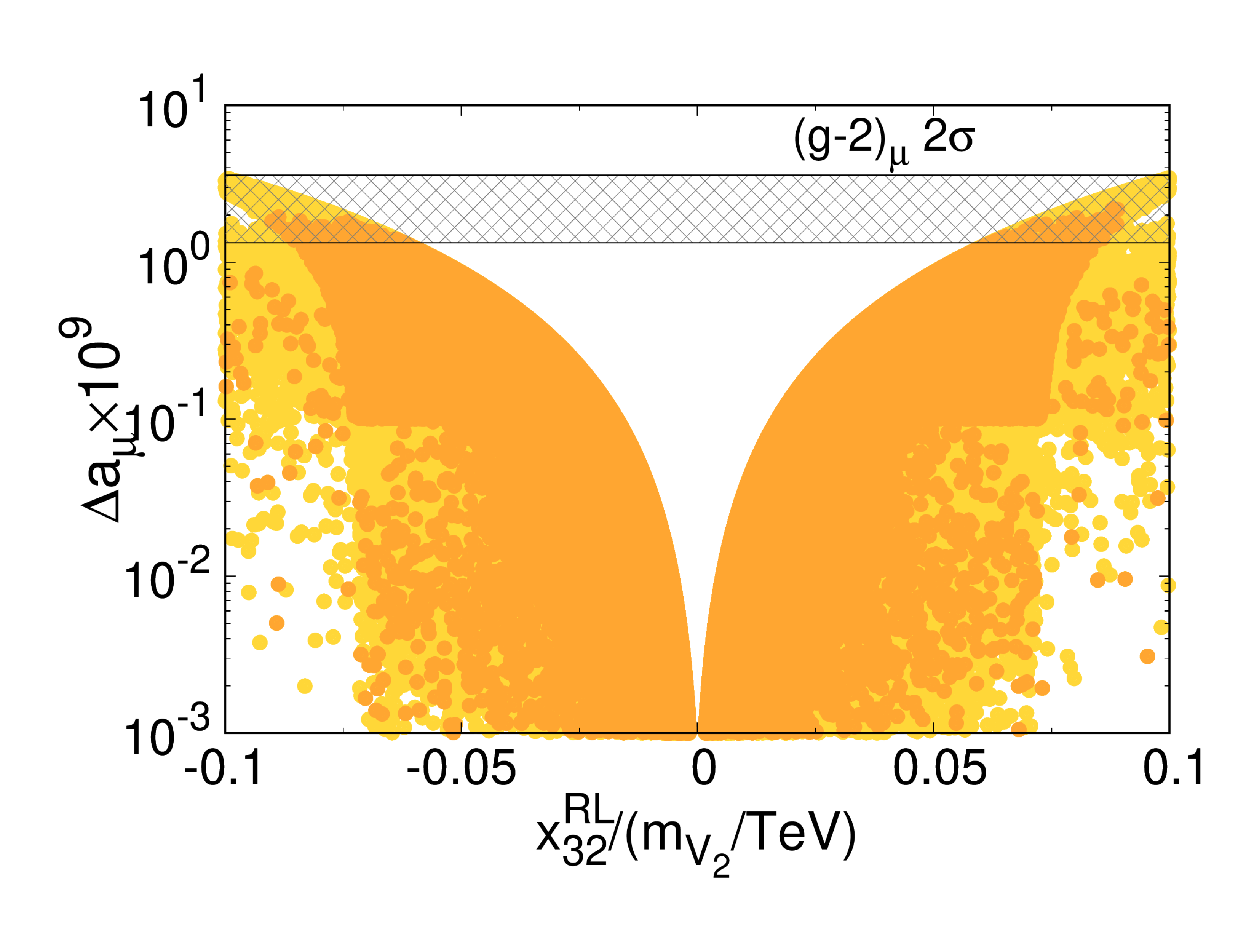}
\includegraphics[height=1.1in,angle=0]{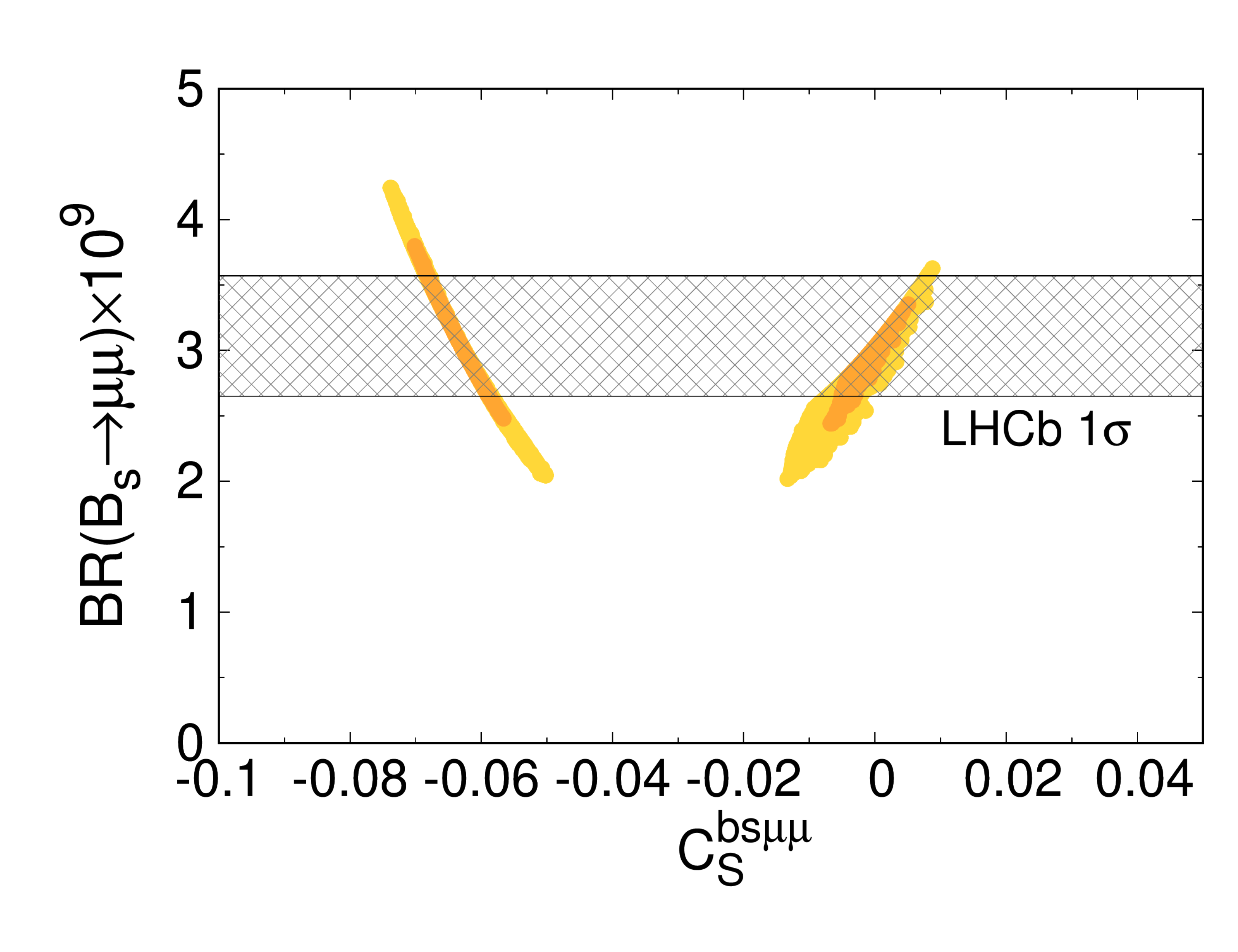}
\includegraphics[height=1.1in,angle=0]{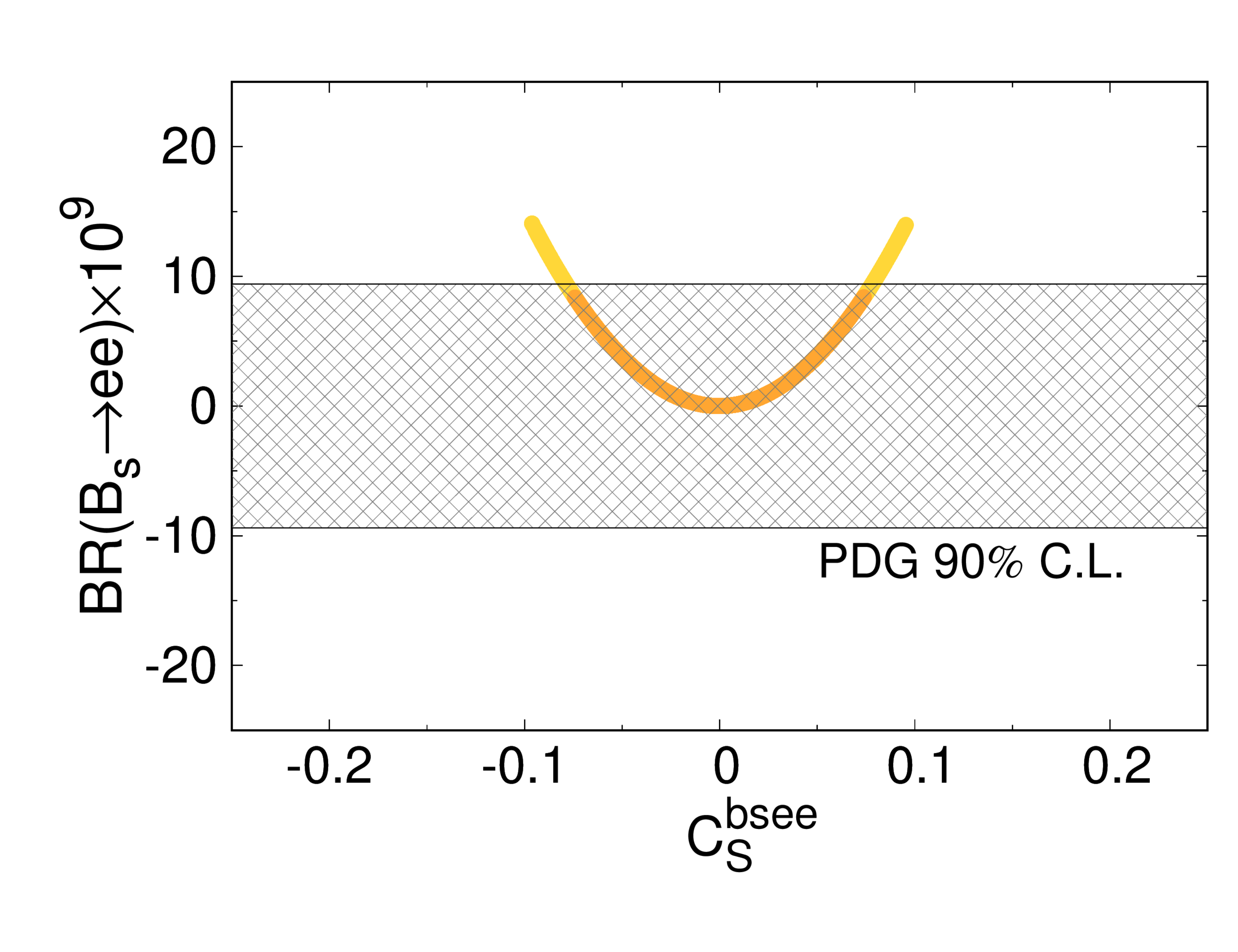}
\caption{\small \label{fig:scan1}
{\bf Scan}: 
The chi-square distribution includes $R_{K,K^*}$, $R_{D,D^*}$, $B_s \to \ell^+ \ell^-$, 
and $\ell_j\to \ell_i \gamma$ data.  
The best-fit point, {\bf BP-1}, gives $\chi^2|_{R_K+R_D+B_s+{\rm LFV}}=22.48$ with pull
3.8$\sigma$ comparing to the SM, $\chi^2|_{R_K+R_D+B_s+{\rm LFV}}=36.92$.
}
\end{figure}

\begin{table}[t]
\caption{\small \label{tab:BP}
Benchmark points with $m_{V_2}=2~{\rm TeV}$.
SM gives $\chi^2|_{R_K+R_D+B_s+{\rm LFV}}=36.92$, 
$\chi^2|_{R_K+R_D+{\rm LFV}+a_\mu+a^{\rm LKB}_e}=57.58$, 
and $\chi^2|_{R_K+R_D+{\rm LFV}+B_s+a_\mu+a^{\rm B}_e}=61.00$.}
\begin{adjustbox}{width=\textwidth}
\begin{tabular}{c|c|c|c}
\hline
\hline
    & {\bf BP-1} & {\bf BP-2} & {\bf BP-3}   \\
\hline
\hline
$X^{\rm LR}_{21}$     
            & -0.445 & -17.85 & -0.0115   \\
$X^{\rm LR}_{31}$     
            & 0.0271 & $6.44\times 10^{-4}$ & 0.927    \\
$X^{\rm RL}_{31}$     
            & $1.58\times 10^{-4}$ & $-2.48\times 10^{-8}$ & -0.0145    \\
$X^{\rm LR}_{22}$     
            & -0.0245 & -$7.18\times 10^{-5}$ & -0.0606   \\
$X^{\rm LR}_{32}$     
            & $-2.01\times 10^{-3}$ & 6.820 & $1.11\times 10^{-5}$   \\
$X^{\rm RL}_{32}$     
            & $-8.25\times 10^{-3}$ & -0.183 & $-1.19\times 10^{-5}$   \\
$X^{\rm LR}_{23}$     
            & -0.367 & $-2.35\times 10^{-3}$ & 0.990   \\
$X^{\rm RL}_{33}$     
            & 0.217 & $-1.87\times 10^{-4}$ & -0.0252   \\
\hline
$C^{bs\mu\mu}_9=C^{bs\mu\mu}_{10}$ 
            & $7.77\times 10^{-3}$ & $-7.73\times 10^{-3}$  & $-1.06\times 10^{-4}$   \\
$C^{bsee}_9=C^{bsee}_{10}$ 
            & -1.90 & -1.82  & -1.68  \\
$C^\ell_{S_R}$ 
        & 0.0276 & $-1.55\times 10^{-4}$  & 0.0141  \\
$C^{bs\mu\mu}_S=-C^{bs\mu\mu}_P$ & -0.0638 & $-4.14\times 10^{-3}$ & $-2.27\times 10^{-4}$ \\
$C^{bsee}_S=-C^{bsee}_P$ & 0.0222 & $-1.40\times 10^{-4}$ & -0.053  \\
\hline
$\Delta a^{V_2}_\mu$ & $-2.11\times 10^{-13}$ & $2.12\times 10^{-9}$ & $-5.19\times 10^{-13}$   \\
$\Delta a^{V_2}_e$ & $-8.02\times 10^{-16}$ & $-1.05\times 10^{-12}$ & $4.52\times 10^{-13}$  \\
\hline
${\rm Br}(\tau \to \mu \gamma)$  & $3.36\times 10^{-11}$ & $7.94\times 10^{-12}$ & $2.04\times 10^{-9}$ \\
${\rm Br}(\tau \to e \gamma)$  & $1.86\times 10^{-8}$ & $1.01\times 10^{-9}$ & $3.65\times 10^{-9}$  \\
${\rm Br}(\mu \to e \gamma)$  & $1.27\times 10^{-13}$ & $6.37\times 10^{-14}$  & $8.68\times 10^{-14}$ \\
\hline
$\chi^2|_{R_K}$ & 21.34 & 22.15 & 21.63  \\
$\chi^2|_{R_D}$ & 0.001 & 1.09 & 0.245  \\
$\chi^2|_{B_s \to \mu\mu}$ & 0.002 & 0.567  & 0.0408   \\
$\chi^2|_{B_s \to ee}$ & 0.018 & 0.000 & 0.548  \\
$\chi^2|_{\rm LFV}$ & 1.12 & 0.065 & 0.156  \\
$\chi^2|_{a_\mu}$ & 18.10 & 0.434 & 18.11  \\
$\chi^2|_{a^{\rm LKB}_e}$ & 2.57 & 26.12 & 0.009 \\
$\chi^2|_{a^{\rm B}_e}$ & 5.96 & 0.232 & 13.69  \\
\hline
$\chi^2|_{R_K+R_D+B_s+{\rm LFV}}$ & 22.48 & 23.88  & 22.62  \\
$\chi^2|_{R_K+R_D+B_s+{\rm LFV}+a_\mu+a^{\rm LKB}_e}$ & 43.15 & 50.43 & 40.74   \\
$\chi^2|_{R_K+R_D+B_s+{\rm LFV}+a_\mu+a^{\rm B}_e}$ & 46.54 & 24.54  & 54.42  \\
$\chi^2|_{R_K+R_D+B_s+{\rm LFV}+a^{\rm LKB}_e}$ & 25.05 & 50.00  & 22.63 \\
$\chi^2|_{R_K+R_D+B_s+{\rm LFV}+a^{\rm B}_e}$ & 28.44 & 24.11  & 36.31 \\
\hline
\hline
\end{tabular}
\end{adjustbox}
\end{table}

We perform the global chi-square fit to the observables,
including $R_{K,K^*}$, $R_{D,D^*}$, $B_s \to \ell^+\ell^-$, and $\ell_i \to \ell_j \gamma$,  
by scanning the couplings of LQ $V_2$ with the scanning ranges of various couplings:
\begin{eqnarray}
\text{\bf Scan}:~~
-20 & \leq & X^{\rm LR}_{21} \leq 20\,,~~
0  \leq  X^{\rm LR}_{31} \leq \sqrt{4 \pi}\,,~~ 
-1  \leq  X^{\rm RL}_{31} \leq 1\,, \nonumber \\
-\sqrt{4 \pi} & \leq & X^{\rm LR}_{22} \leq \sqrt{4 \pi}\,,~~
-2\sqrt{4 \pi}  \leq  X^{\rm LR}_{32} \leq 2\sqrt{4 \pi}\,,~~
-1  \leq  X^{\rm RL}_{32} \leq 1\,, \nonumber \\
-1 & \leq & X^{\rm LR}_{23} \leq 1\,,~~ 
-2  \leq  X^{\rm RL}_{33} \leq 2\,,~~  
m_{V_2}=2~{\rm TeV}\,.
\end{eqnarray}
Since the theoretical uncertainties are still ambiguous among $\Delta a_{e,\mu}$,
we have not included the observables in the global fit at this stage,
but rather treat them as posterior predictions.
Due to the facts that all the observables and Wilson coefficients are originated 
from products of two couplings, 
such that only the relative sign between couplings would be revealed
from the chi-square fitting,
we scanned the positive value of $X^{\rm LR}_{31}$ and both signs for the other couplings.
Note that the $X^{\rm LR}_{21}$ and $X^{\rm LR, RL}_{31}$
($X^{\rm LR}_{22}$ and $X^{\rm LR,RL}_{32}$) 
are related to $C^{bsee}_{9,10}$
($C^{bs\mu\mu}_{9,10}$).
The $X^{\rm LR}_{23}$ combining with $X^{\rm RL}_{3\ell}$ dominately
contributes to $C^\ell_{S_R}$,
while other combinations are suppressed due to the off-diagonal elements of CKM matrix.
Since the flavors of neutrino are indistinguishable in the process $b\to c \tau \nu_\ell$, 
we sum over the neutrino flavors.
The $\Delta a_e$ ($\Delta a_\mu$) are generated 
through a pair of bottom-quark and $V_2$ running in the loop, 
since the heavy $b$-quark mass enhances flipping the chiralities of external muons,
thereby they are strongly correlated to $X^{\rm LR,RL}_{31}$ ($X^{\rm LR,RL}_{32}$).
We adopted the two-dimensional chi-square statistics, i.e
$\Delta \chi^2\leq$ 2.30 (6.18) corresponds to 1-$\sigma$ (2-$\sigma$) regions. 
The chi-square distributions are shown in Fig.~\ref{fig:scan1}, 
based on projections of two selected parameters whereas marginalizing the others.

Among the scanning points using the data
$R_{K,K^*}$, $R_{D,D^*}$, $B_s \to \ell^+\ell^-$, and $\ell_i \to \ell_j \gamma$,
we further select three benchmark points, 
which yield minima of chi-square with respect to various groups of observables,
and more details are listed in Table~\ref{tab:BP}: 
\begin{itemize}
\item {\bf BP-1}: the best fit to $R_{K,K^*}$, $R_{D,D^*}$, $B_s \to \ell^+ \ell^-$, 
and LFV data
and gives $\chi^2|_{R_K+R_D+B_s+{\rm LFV}}=22.48$ with a pull 3.8$\sigma$ 
comparing to the SM, $\chi^2|_{R_K+R_D+B_s+{\rm LFV}}=36.92$.

\item {\bf BP-2}:  gives $\chi^2|_{R_K+R_D+B_s+{\rm LFV}+a_\mu+a^{\rm B}_e}=24.54$, and
thus provides a simultaneous solution to $(g-2)_\mu$ and $(g-2)^{\rm B}_e$.

\item {\bf BP-3}: gives $\chi^2|_{R_K+R_D+B_s+{\rm LFV}+a^{\rm LKB}_e}=22.63$,
thus provides an explanation for $(g-2)^{\rm LKB}_e$.
\end{itemize}

Because the iso-doublet vector LQ $V_2$ couples to the right-handed lepton 
and thus induces $C^{bsee}_9=C^{bsee}_{10}$,
the $(C^{bsee}_9,C^{bsee}_{10})$ panel in Fig.~\ref{fig:scan1}.
The {\bf BP-1} in Table~\ref{tab:BP} 
shows that the $C^{bsee}_9=C^{bsee}_{10}=-1.90$ from $V_2$ provides the best solution 
for $R_{K,K^*}$ anomaly,
which increases $b\to s e^+ e^-$ to reduce the values of $R_{K,K^*}$.
It implies the correlation
\begin{eqnarray}
\label{eq:RK_limit}
\frac{X^{\rm LR}_{31}(X^{\rm LR}_{21})^*}{m^2_{V_2}}\simeq \frac{-0.0030}{(1~{\rm TeV})^2}
\end{eqnarray}
and is shown in the $(-X^{\rm LR}_{21},X^{\rm LR}_{31})$ panel in Fig.~\ref{fig:scan1}.

The $(X^{\rm LR}_{23},X^{\rm RL}_{33})$ panel in Fig.~\ref{fig:scan1} 
indicates mild correlation between $(X^{\rm LR}_{23}$ and $X^{\rm RL}_{33})$,
that came from the $R_{D,D^*}$ observables and thus Wilson coefficient $C^\ell_{S_R}$.
According to the preferred value of $C^\ell_{S_R}$ from Table~\ref{table-data},
Eq.(\ref{eq:ClSR}) gives
\begin{eqnarray}
\frac{X^{\rm RL}_{33}(X^{\rm LR}_{23})^*}{m^2_{V_2}}\simeq \frac{-0.019}{(1~{\rm TeV})^2}\,,
\end{eqnarray}
where the minus sign is originated from the $V_{cd}\simeq -0.041$.
Since $C^\ell_{S_R}=0$ is still compatible with observation within 1-$\sigma$ ($\Delta \chi^2\leq 2.3$),
the chi-square regions in $(X^{\rm LR}_{23},X^{\rm RL}_{33})$ panel connected. 

The $B_s \to \mu^+ \mu^-$ observable dictates the chi-square regions 
in $(C^{bs\mu\mu}_S,C^{bs\mu\mu}_P)$ panel, 
and there are two solutions, 
$C^{bs\mu\mu}_S=-C^{bs\mu\mu}_P\simeq -0.064$ and $\simeq 0.00$ correspond respectively to non-SM and SM solutions,
which also exhibit in the $(X^{\rm LR}_{22},X^{\rm RL}_{32})$ 
and $(\Delta a_\mu,X^{\rm LR}_{22})$ panels.
The non-SM and SM solutions require
\begin{eqnarray}
\frac{X^{\rm RL}_{32}(X^{\rm LR}_{22})^*}{m^2_{V_2}}\simeq \frac{-5.1\times 10^{-5}}{(1~{\rm TeV})^2}\,,~\text{and}~~~~
\frac{|X^{\rm RL}_{32}(X^{\rm LR}_{22})^*|}{m^2_{V_2}}\lesssim \frac{10^{-5}}{(1~{\rm TeV})^2}\,,
\end{eqnarray}
respectively.
On the other hand, the  $B_s \to e^+ e^-$ observable, which is consistent with SM,
set limit on
\begin{eqnarray}
\label{eq:bsee_limit}
\frac{|X^{\rm RL}_{31}(X^{\rm LR}_{21})^*|}{m^2_{V_2}}\lesssim \frac{10^{-4}}{(1~{\rm TeV})^2}\,,
\end{eqnarray}
explaining the region in $(X^{\rm LR}_{21},X^{\rm RL}_{31})$ panel in Fig.~\ref{fig:scan1}.

{\bf BP-1} fits well to $R_{K,K^*}$, $R_{D,D^*}$, $B_s \to \ell^+ \ell^-$, 
and LFV observables, however does not induce sizable $\Delta a_\mu$ or $\Delta a_e$.
{\bf BP-2} intriguingly provides simultaneous solutions to
$\Delta a_\mu$ and $\Delta a^{\rm B}_e$,
as well as a decent fit to B-physics observables.
The severe restriction from $\mu \to e \gamma$ is the main difficulty to a generate sizable 
$\Delta a_\mu$. 
Specifically, the former is related to the linear combinations of
$X^{\rm LR}_{31}X^{\rm RL}_{32}$ 
and $X^{\rm LR}_{32}X^{\rm RL}_{31}$, 
while the latter is related to the product $X^{\rm LR}_{32}X^{\rm RL}_{32}$.
Therefore, suppressing the $X^{\rm LR,RL}_{31}$ to avoid the  $\mu \to e \gamma$ constraint 
may help to unleash a large enough $\Delta a_\mu$.
According to Eq.(\ref{eq:RK_limit}) and (\ref{eq:bsee_limit}), apparently, 
the above scenario can be achieved by a large enough $|X^{\rm LR}_{21}|$,
and this is main result of {\bf BP-2}.
Unfortunately, the minimum requirement of $|X^{\rm LR}_{21}|$ 
is significantly above the perturbative limit, $\sqrt{4 \pi}$.
For example, it needs
$X^{\rm LR}_{21}\lesssim -8$ from $(X^{\rm LR}_{21},\Delta a_\mu)$ panel
in order to start overlapping with 2-$\sigma$ region of $(g-2)_\mu$, and 
it is $X^{\rm LR}_{21}=-17.85$ for {\bf BP-2}.
Finally, {\bf BP-3} provides an alternative solution 
for $\Delta a^{\rm LKB}_e$ and B anomalies,
meanwhile is consistent with LFV and B-physics limits.

\begin{figure}[t!]
  \centering
    \includegraphics[width=4.8in,angle=0]{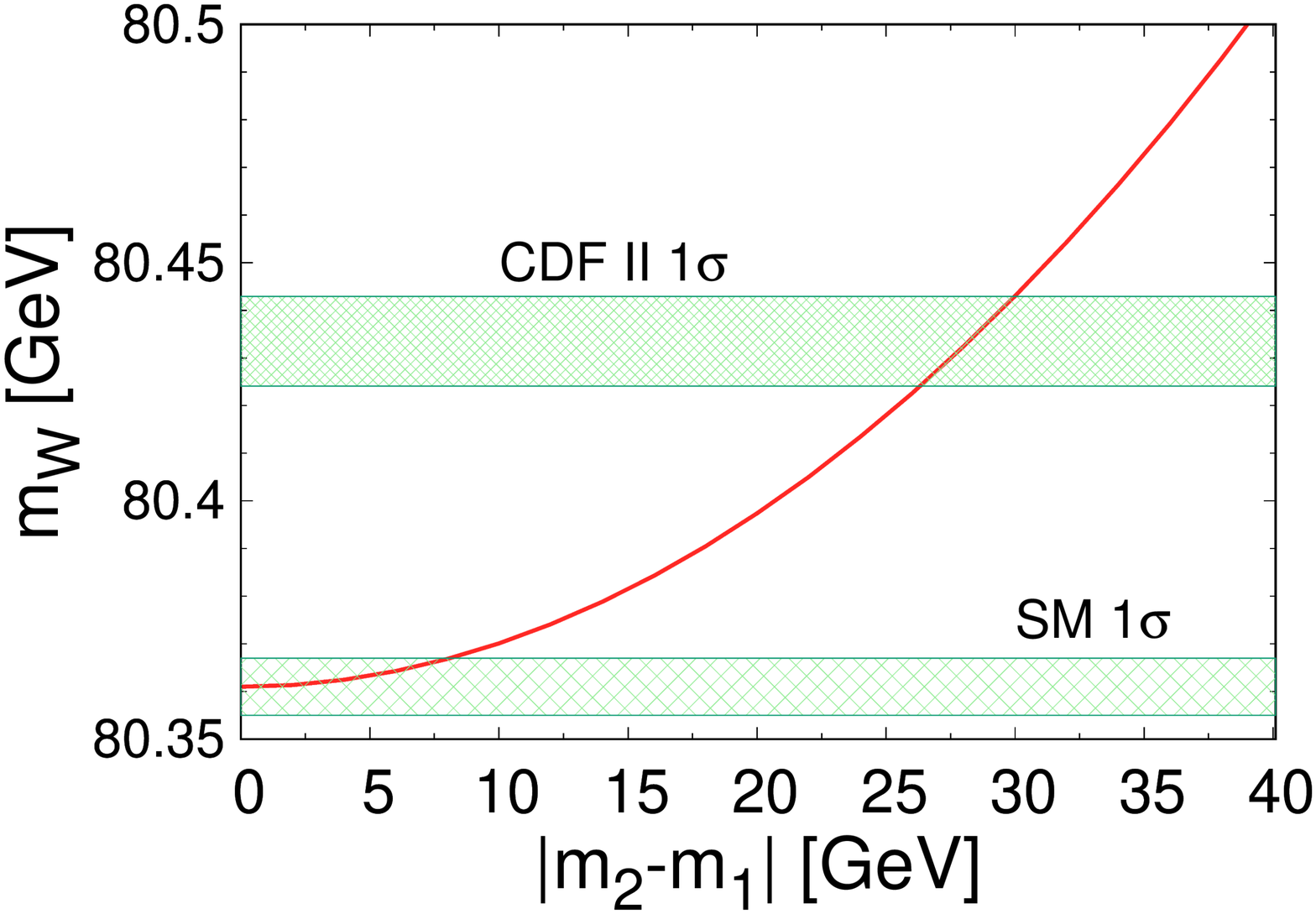}
    \caption{\small \label{fig:mw}
      The resulting $W$-boson mass due to the mass splitting between the upper and
      lower isospin component of the vector LQ $V_2$ around 2 TeV. 
      Note that the lower band in green
      is the SM prediction while the upper band in the latest CDF measurement.
       }
\end{figure}

The raising of the $W$-boson mass due to the mass splitting of the iso-doublet $V_2$ is given
in Eq.~(\ref{raise}). The raise only depends on the absolute of the mass difference and
is shown in Fig.~\ref{fig:mw}.  It can be seen that a mass splitting of 
$25\text{-}30$ GeV
provides a viable solution to the $W$-boson anomaly. Such a mass splitting
corresponds to $1.25\% - 1.5\%$ of a 2 TeV LQ. 

\section{Conclusions}

Most works in literature  on solving the $R_{K,K^*}$ anomaly rely on reducing
$b \to s \mu^+ \mu^-$ with the LQ couplings to the left-hand muon.  Nevertheless,
it remains an almost equally viable solution of increasing $b \to s e^+ e^-$.
Here in this work, we have attempted to use the iso-doublet vector LQ $V_2$
that couples to the right-handed electron to increase $b \to s e^+ e^-$ and found
parameter space to explain the $R_{K,K^*}$. Simultaneously, it can also explain
the $R_{D,D^*}$ and consistent with the $B_s$ decays.

We have also investigated the possibility of explaining the muon anomalous magnetic
moment. Such a possibility is severely constrained by the leptonic radiative decay
$\ell_i \to \ell_j \gamma$.   We have successfully found some parameter space points
that can explain all $R_{K,K^*}$, $R_{D,D^*}$, $B_s$ decays, LFV, and $\Delta a_\mu$
and $\Delta a^{\rm B}_e$ (see {\bf BP-2}), though one of the couplings is close to or
larger than the perturbative limit.

Furthermore, the iso-doublet vector leptoquark $V_2$ naturally explains the $W$-boson
anomaly with a mass splitting of order $25\text{-}30$ GeV between the isospin components.

\section*{Acknowledgment}
We specially thank to Wolfgang Altmannshofer for useful discussion on renormalization of
Wilson coefficients, and to Chih-Ting Lu for discussion on $W$-boson anomaly.
K.C. also thanks Wai-Yee and Gum-see Keung for their great hospitality.
The research was supported in part by the Ministry of Sciences and Technology with
grant number MoST-110-2112-M-007-017-MY3 and MoST-111-2112-M-007-012-MY3.

{\it Note Added:} We came across a number of works in attempt to explain the
$W$-boson anomaly \cite{Fan:2022dck,Zhu:2022tpr,deBlas:2022hdk,Yang:2022gvz,Athron:2022qpo,Strumia:2022qkt,Arias-Aragon:2022ats,Athron:2022isz,Heckman:2022the,Sakurai:2022hwh,Lu:2022bgw}.

\bigskip


\end{document}